\documentclass[aps,pre,10pt,showpacs,superscriptaddress,floatfix,reprint]{revtex4-1}
\usepackage{amsmath}    
\usepackage{graphicx}   
\usepackage{verbatim}   
\usepackage{color}      
\usepackage{subfigure}  
\usepackage{hyperref}   
\usepackage{pstricks}

\usepackage{epsfig}
\usepackage{amsmath}
\usepackage{amssymb}

\usepackage{graphicx}

\definecolor{BoxCol}{rgb}{0.9,0.9,1}

\definecolor{SectionCol}{rgb}{0,0,0.5}

\def\C{\mathrm{k}}

\renewcommand{\Re}[1]{\mathop{\rm Re}\nolimits\left\{ #1\right\}}
\renewcommand{\Im}[1]{\mathop{\rm Im}\nolimits\left\{ #1\right\}}
\renewcommand{\min}[1]{\mathop{\rm min}\nolimits\left\{ #1\right\}}
\renewcommand{\max}[1]{\mathop{\rm max}\nolimits\left\{ #1\right\}}

\def\mkm{{{$\mu$m}}}
\def\bs{\boldsymbol}

\newcommand{\RPA}{{\mathrm{RPA}}}

\newcommand{\eps}{{\varepsilon}}

\newcommand{\rf}[1]{~(\ref{#1})}
\def\dst{\displaystyle}
\def\sst{\scriptstyle}

\newcommand\Tr{\mathop{\rm Tr}\nolimits}
\newcommand\Res{\mathop{\rm Res}\nolimits}

\newcommand\Ai{\mathop{\rm Ai}\nolimits}
\newcommand\Bi{\mathop{\rm Bi}\nolimits}

\newcommand\Fermi{\mathrm{F}}

\newcommand\FF[1]{~Fig.\,\ref{f:#1}}

\definecolor{emphc}{rgb}{1,1,0.44}

\renewcommand{\emph}[1]{{\textbf{#1}}}

\def\nuef{\nu_\mathrm{eff}}
\def\nuefmax{\nu_\mathrm{eff,max}}

\newcommand{\chir}{\chi_{_{\!\!\sst{Z_1}}}}
\newcommand{\chii}{\chi_{_{\!\!\sst{Z_2}}}}
\newcommand{\chiii}{\chi_{_{\!\!\sst{Z_3}}}}

\newcommand{\alphaz}{\alpha_{_{\!\!\sst{Z}}}}
\newcommand{\betaz}{\beta_{_{\!\!\sst{Z}}}}
\newcommand{\betazz}{\widetilde{\beta}_{_{\!\!\sst{Z}}}}
\newcommand{\fF}{f_{\sst\mathcal{F}}}

\def\Zef{Z_{\ast}}
\def\nuplo{\nu_{pl,0}}

\def\nuplo{\nu_\mathrm{pl,0}}

\newcommand{\maxxi}[1]{{\raise-2.mm\hbox{$\textstyle  {\rm max}
 \atop {\raise1.7mm\hbox{$\scriptscriptstyle \xi$}}$}}\! \left\{ #1\right\}}

\def\toinfty{\hbox{\space \raise-1.5mm\hbox{ $\textstyle  \longrightarrow
 \atop {\raise1mm\hbox{$\scriptstyle \r\to\infty$}}$} \space}}

\def\eps{\varepsilon}
\def\C{\mathrm{k}}
\def\K{\mathcal{K}}

\def\ef{\epsilon_{\Fermi}}
\def\emu{\epsilon_\mu}

\def\M{\bs{M}}

\def\Z{\textsl{Z}}

\def\k{{\tilde{k}}}
\def\tnu{{\tilde{\nu}}}
\def\w{{\tilde{\omega}}}
\def\R{{\tilde{R}}}

\def\CL{C_{\Lambda,0}}
\def\CCL{C_{\Lambda,1}}

\def\epsrpa{\varepsilon_{\rm RPA}}

\def\C{\mathrm{k}}

\def\debroil{{\!\lefteqn{\lambda}\;\bar{}\:}}
\def\Cf{\mathfrak{C}}
\def\Af{\mathfrak{A}}
\def\Nf{\mathfrak{N}}
\def\wstar{\omega^*}
\def\I{\mathcal{I}}
\def\F{\mathcal{F}}

\begin{document}

\title{Optical conductivity of warm dense matter in wide frequency range within
 quantum statistical and  kinetic approach}
 
\author{M. Veysman}
\affiliation{Joint Institute for High Temperatures (JIHT) RAS, Izhorskaya 13/19, Moscow 125412, Russia}


\author{G. R\"opke}
\affiliation{Universit\"at Rostock, Institut f\"ur Physik, 18051 Rostock, Germany}

\author{M. Winkel}
\affiliation{Darmstadt}

\author{H. Reinholz}
\affiliation{Universit\"at Rostock, Institut f\"ur Physik, 18051 Rostock, Germany}
\affiliation{The University of Western Australia, School of Physics, Crawley, WA 6009, Australia}

\email{heidi.reinholz@uni-rostock.de}

\date{\today}


\begin{abstract}

Fundamental properties of warm dense matter are described by the
dielectric function, which
gives access to the frequency-dependent electrical conductivity,
absorption, emission and scattering of
radiation, charged particles stopping and further macroscopic
properties. Different approaches to the dielectric function and
the related dynamical collision frequency  are
compared in a wide frequency range.
The high-frequency limit describing inverse bremsstrahlung and the
low-frequency limit of
the dc conductivity are considered. Sum rules and Kramers-Kronig
relation are checked for the 
generalized linear response theory and the standard approach
following kinetic theory. The results are discussed in application to
aluminum, xenon and argon
plasmas.

\end{abstract}

\maketitle

\setlength{\parskip}{2mm}

\section{Introduction}

Interaction of laser radiation with matter is utilized for many modern applications, like creation of sources of high energy particles and short wavelength radiation~\cite{Pugachev15PPR,Pugachev15LPB,Kostenko16PoP,Kostenko13QEl}.
Under irradiation of solid targets with intense laser pulses, matter undergoes transformations from a cold solid up to
hot  and dense plasma (warm dense matter, WDM) and further to weakly coupled plasmas with properties rapidly  varying in time and space.
Therefore, for a correct description of laser-matter interaction in very different regions of parameter values for mass density,  electron and ion  temperatures 
one needs wide-range models for the optical properties of  WDM, which are determined by the permittivity or dielectric function (DF) $\eps({\bs{k}}, \omega)$. 

Knowing wide-range expressions for $\eps({\bs{k}}, \omega)$ and also transport coefficients and equations of state, one can determine space and time dependencies of laser heated matter by means of hydrodynamic codes such as Lasnex \cite{LASNEX77}, Medusa \cite{Dajaoui92JPB}, Multi-fs \cite{Eidmann00PRE}, or the code of the JIHT group \cite{Veysman08JPB,Povarnitsyn12AppSS,Povarnitsyn13LPB,Andreev15LPB}. Primarily,  those codes use semi-empirical models for $\eps({\bs{k}}, \omega)$ and a corresponding effective electron-ion (electron-phonon) collision frequency~\cite{Agranat07JETPHL,Veysman08JPB}, which are derived from kinetic equations  and give known limits for the case of weakly-coupled plasmas, hand-book values for cold solid, and are based on physically reasonable estimates and experimental data in the intermediate region~\cite{Eidmann00PRE,Povarnitsyn12AppSS,Veysman08JPB}. The elaboration of a systematic many-particle approach that covers distinct regions like cold metals and hot strongly coupled plasmas is a challenging problem in non-equilibrium statistical physics.

Besides this, the advantage of such an approach is the description of laser interaction with matter for a wider range of laser parameters, from infrared to X-ray wavelengths. This is requested taking in mind recent achievements in the construction of powerfull laser systems operating in ultraviolet and X-ray wavelengths~\cite{Schmuser2014FEL,Sekutowicz15PRSTAB,Reagan14PRA}.  Note that local thermodynamic equilibrium in WDM is established on the fs scale after the excitation by laser irradiation, see e.g.\cite{Sperling15}, 
and will be assumed for the following considerations. 

The most strict many particle approach for calculating the permittivity of WDM consists of a quantum statistical (QS) description  of the reaction of the system to 
external perturbations, see~\cite{ZMR97} (often also called  ``Zubarev approach``).
Within the QS approach, both  fundamental theoretical approaches can be derived, as it was demonstrated in a recent 
paper~\cite{Reinholz12PRE}:  the linear response theory (LRT)~\cite{ReinholzADP,Roepke98PRE} follows from the QS formalism if one 
chooses moments of the particle distribution function as relevant observables, and the kinetic approach follows if density 
fluctuations are chosen as sets of relevant observables. In turn, the kinetic approach can be realized on the basis of quantum 
kinetic equations~\cite{Kull01PoP}, or, alternatively, using classical kinetic theory (KT)~\cite{LL_PhysKin} and the concept of 
cross-sections, that leads to the formulation of kinetic equations with Boltzmann or Fokker-Planck collision integrals. 
In the most simple 
form, when one can disregard electron-electron collisions, the electron-ion collision integral can be written in relaxation time 
approximation, which leads to  simple expressions for the permittivity, which are widely used in hydrodynamic 
codes~\cite{Andreev15LPB,AndreevTVT03,VeysmanPoP06,Eidmann00PRE}.  

Another approach using  a classical method of moments which satisfies the sum rules \cite{Tkachenko14} is a promising alternative to derive analytical approximations but will not be further considered here.

Following LRT, transport coefficients and expressions for inverse bremstrahlung absorption are expressed by equilibrium correlation functions which can be calculated with the help of the Green functions technique in a systematic way. This procedure takes consistently into account  many-particle effects, such as electron and ion correlations and dynamical screening, and also effects of strong collisions relevant for large-angle scattering~\cite{Reinholz00PRE,ReinholzADP}. Account of these effects can be essential for  studies of optical properties of  WDM, i.e., at temperatures of the order of $T\sim 0.1\div 10^2$ eV and densities up to the order of solid ones \cite{Sperling15}.
An alternative to the perturbative treatment using Green functions leading to analytical expressions is the direct evaluation of the equilibrium correlation functions using molecular dynamics (MD) simulations of  ions combined with the density functional theory (DFT) for the electrons as denoted by the Kubo-Greenwood approach, see~\cite{Reinholz15PRE,Knyazev14PoP,Norman13CPP,Zhilyaev13DokPh}. The reliability of results obtained by means of perturbative (using Green functions) approaches was confirmed recently by comparison  with numerical MD-simulations~\cite{Morozov04PRE,Morozov05PRE,Reinholz15PRE}.

A consistent treatment of strong-coupling effects and the frequency dependence of the dynamical conductivity~\cite{Reinholz00PRE,Reinholz12PRE} are the strength of the LRT approach. On the other hand, the respective expressions can be rather cumbersome and therefore difficult for implementation in hydrodynamic codes. This is why the elaboration of approximate semi-empirical formulas and interpolation models is of great interest, especially taking in mind the necessity to make connection with experimental data and MD simulations in the region of WDM parameters, where one can't extract small parameters for the theory to be build. 

The principal ideas of the LRT approach will be summarized in Sect. \ref{sec2}. In particular, a generalized screening parameter taking into account dynamical screening effects is proposed. The KT following the solution of the Fokker-Planck-equations is described in Sect. \ref{sec:relaxtime}. Results are discussed in Sect. \ref{sec4}. A comparison will be made between calculations of the DF using LRT  with results obtained from the semi-empirical model~\cite{Agranat07JETPHL,Veysman08JPB,Povarnitsyn12AppSS} based on KT and with other models utilizing the concept of relaxation time approximation and Coulomb logarithm~\cite{Stygar2002PRE,Skupsky87PRA}.  In addition, calculations of an effective frequency for electron-ion collisions on the basis of quantum~\cite{Reinholz12PRE} and classical~\cite{LL_PhysKin} kinetic equations are compared. The influence of plasma inhomogenities and inter-band transitions will be discussed in Sect. \ref{sec5} and investigated considering the reflectivity of shock compressed plasmas.

\section{Dielectric function from quantum statistical approach using LRT} \label{sec2}

 Warm dense matter can be described as system of interacting particles, electrons and ions. 
In contrast to a first principle approach that treats the electrons and atomic nuclei as constituents, 
we consider  ions and free electrons. The latter are unbound electrons or electrons in conduction band  with  
free electron density $n$ and temperature  $T$. We use energy units, thus setting $k_{B}=1$.
In general, there are  ions with different ionization and excitation stages. 
For simplicity we consider the ion component in terms of an average charge number $Z$ with the particle 
density $n_{\rm i}=n/Z$ due to charge neutrality. 
The ion temperature is denoted as $T_{\rm i}$.   
The CGS system of units is used in the following, thus replacing $e^2/(4 \pi \epsilon_0)$ 
in previous papers (e.g. Refs. \cite{Reinholz00PRE,Reinholz12PRE}) by $e^2$. 

The ions are treated in adiabatic approximation~\cite{Karakhtanov13CTPP} 
via the static ion structure factor,
 $S_{\rm ii}({\bs{k}})= n_{\rm i}^{-1}\sum\nolimits_{i,j} \langle \exp[i {\bs{k}} \cdot ({\bs{R}}_i - {\bs{R}}_j)]\rangle$
 which describes static correlations of ions at positions  ${\bs{R}}_j$ such as lattice formation (ions in a 
unit volume are considered; brackets $\langle..\rangle$ denote statistical average).  

In the liquid phase, the ion structure factor also has to be considered.
For the interaction of electrons with collective excitations of the ion lattice, i.e. phonons, one should consider time dependent positions ${\bs{R}}_j(t)$ leading to the dynamical ionic structure factor  $S_{\rm ii}({\bs{k}}, \omega)= n_{\rm i}^{-1} \sum\nolimits_{i,j} \int dt \exp[-i \omega t] \langle \exp[i {\bs{k}} \cdot ({\bs{R}}_i(t)- {\bs{R}}_j)]\rangle$. 

In the  below derivation within LRT and KT, the static structure factor is taken to be $S_{\rm ii}({\bs{k}})=1$ for simplicity (non-correlated ions). The influence of ion correlations is considered later in Secs.~\ref{s:Sii} and~\ref{s:results}. 

We express the DF in terms of 
equilibrium correlation functions. For the Hamiltonian of the electron-ion system we consider the electronic degrees of freedom only 
\begin{eqnarray}
\label{H}
&& H= \sum_p E_p  \hat a_p^{\dagger}\hat a_p + \sum_{pk}V_{\rm ei}(k)  
\hat a_{p+k}^{\dagger}\hat a_p
\nonumber \\&& +\frac{1}{2} \sum_{p_1 p_2k} V_{\rm ee}(k) 
\hat a_{p_1+k}^{\dagger} \hat a_{p_2-k}^{\dagger}\hat a_{p_2}\hat a_{p_1}\,.
\end{eqnarray}
with $E_p=\hbar^2p^2/(2m)$.  Interactions between ions and electrons $V_{\rm ei}(k)=V(k)$ are given by the Coulomb potential 
$ 
V(k) = -Z v(k)  \quad \mbox{and } \; V_{\rm ee}(k)=v(k)=4\pi  e^2 /k^2
$ 
 is the potential of the e-e interaction. 

Note that in the general case pseudo-potentials $V_{\rm ei}^{\rm ps}(k)$  may have to be considered. 
They reflect the fact that the fundamental Coulomb interaction between charged particles is modified 
if we introduce quasiparticles such as band electrons in the lattice formed by ions. 
Besides this, the structure of complex ions can also influence on their interaction with free electrons 
via interaction of bound and free electrons. This is discussed in Secs.~\ref{s:Sii} and~\ref{s:results}  
at a phenomenological level. In expressions given below for the Born approximation, we have to replace
\begin{equation}
|V_{\rm ei}({{\bs{k}}})|^2= |V_{\rm ei}^{\rm ps}(k)|^2 S_{\rm ii} ({{\bs{k}}}).
\label{Vei}
\end{equation}

Generally speaking, time variation of the current density contains intra-band and inter-band contributions. Consequently, 
the response of the system is determined by 
intra-band  (single band, {\it sb}) as well as inter-band (bound - bound, {\it bb}) scattering mechanisms. 
Below we restrict ourselves to intra-band contributions which are described by a plasma model for WDM. The inter-band contribution to the DF is considered in Secs.~\ref{s:bb} at a more phenomenological level.

In the present  treatment we disregard electron-phonon interactions and Umklapp processes, which is valid at temperatures sufficiently higher than melting temperatures. Also,  temperatures are considered such that the plasma coupling parameter $\Gamma_{\rm ei} = Z e^2 /(R_0 T) \lesssim 1$,
where
\begin{equation}
R_0 = (4\pi n_{\rm i}/3)^{-1/3}
\label{R0}
\end{equation}
is the Wigner-Seitz radius.
 The degeneracy parameter
$\Theta = \ef^{-1}= T/E_{\rm F} = 2mT/\hbar^2 (3\pi^2 n)^{-2/3}$,
where  $E_{\rm F}$ is the Fermi energy, can be arbitrary, if it is not stated otherwise.

Before proceeding further, common expressions for DF and it's connection with polarization and response functions are considered briefly.

\subsection{General expressions for the contribution of electrons to the DF}

In accordance with common theory~\cite{Bogdankevich84}, the permittivity of an isotropic medium is expressed as $\eps_{ij}(\bs{k},\omega)=(\delta_{ij}-k_i k_j/k^2)\eps_\bot(\textbf{k},\omega) +
k_i k_j/k^2 \eps_\|(\bs{k},\omega)$, 
where $\eps_\|, \eps_\bot, \omega $ and $\bs{k}$ are longitudinal, transverse parts of permittivity, frequency of radiation and wave vector, respectively. $k=|\bs{k}|$, indexes $i,j$ denote respective components. From Maxwell equations, taking in mind electric field of polarization charge and density fluctuations, one can find the following equivalent expressions for the longitudinal permittivity $\eps$~\cite{Pines99,ZMR97} (here and below, index ``$\|$``  is omitted for brevity):
\begin{equation}
\eps(\bs{k},\omega)=1-v(k)\Pi(\bs{k},\omega) = \Bigl[1+v(k)\chi(\bs{k},\omega)\Bigr]^{-1},
\label{epsPichi}
\end{equation}
where 
$\Pi(\bs{k},\omega) = \rho({\bs{k}}, \omega)/\phi_{\rm tot} (\bs{k},\omega)$ is the polarization function and 
$\chi(\bs{k},\omega) = \rho({\bs{k}}, \omega)/\phi_{\rm ext} (\bs{k},\omega)$ is the response function.   
$\phi_{\rm tot}=\phi_{\rm pol}+\phi_{\rm ext}$ is the total scalar potential consisting of the external potential  
$\phi_{\rm ext}$ and the potential of polarization charge $\phi_{\rm pol}$, 
$\rho({\bs{k}}, \omega)$ is the Fourier transform of the quantum mechanical average of density fluctuations, 
$\rho({\bs{k}}, t)=\langle \Psi(t)|\rho_{\bs{k}}|\Psi(t)\rangle$, where 
$\rho_{\bs{k}} = \sum_l \! \int\nolimits d^3 r \delta(\bs{r}-\bs{r}_l)
e^{-i\bs{k}\bs{r}}=\sum_l \! e^{-i\bs{kr}_l}$ is the Fourier component of the charge particle density and $\Psi(t)$ the full wave function of the system.
According to Eq.\rf{epsPichi}, the function $\chi(\bs{k},\omega)$ is connected to the
polarization function $\Pi(\bs{k},\omega)$ by the relation
\begin{equation}
\Pi(\bs{k},\omega) = \chi(\bs{k},\omega)/[1+v(k)\chi(\bs{k},\omega)].
\label{Pichi}
\end{equation}
The response function determines the reaction of the system on external perturbations. 

At zeroth order of interaction, the permittivity  
is determined by the polarization function in random phase approximation (RPA)~\cite{ZMR97,Gorobchenko80UFN,Pines99}
\begin{eqnarray}
\Pi_\RPA(\bs{k},\omega) & = &\chi_0(\bs{k},\omega) \\
&= &2\sum\limits_{\bs{p}} \frac{f_{\bs{p}+\bs{k}}-f_{p}}{E_{\bs{p}+\bs{k}}-E_{p}-(\hbar\omega +i\delta)},
\label{PiRPA}
\end{eqnarray}
 where spin summation gives the factor 2,
$f_{\bs{p}} =f_p = [1+e^{(E_{\bs{p}}-\mu)/T}]^{-1}$ is the electron distribution function, 
$\mu$   is the chemical potential of the electrons, and
the limit $\delta \to +0$ is  considered. 

If the interaction between the charged particles is taken into account, the local microscopic field acting on an electron differs from the average macroscopic field. In this case the polarization of the electron gas is different from the sum of the polarization of it's individual particles~\cite{Gorobchenko80UFN} and the RPA polarization function\rf{PiRPA} is replaced by
\begin{equation}
	\label{Pi}
 \Pi (\bs{k}, \omega) = \frac{\Pi_\RPA(\bs{k},\omega)}{1+G(\bs{k},\omega)\Pi_\RPA(\bs{k},
      \omega) v(k)}, 
\end{equation}
 where $G(\bs{k},\omega)$ is  the \textit{local field factor}~\cite{Gorobchenko80UFN,Roepke99PRE,Reinholz00PRE}, which contains all effects due to   charged particle interactions, in particularl, dynamical screening, correlations, and strong collisions.


\subsection{Response function and local field factor in terms of correlation functions}

Using the method of quantum statistical operator~\cite{ZMR97}, within LRT one can show~\cite{Roepke98PRE,RoepkeWierling98PRE,Reinholz00PRE} that the response function is
\begin{equation}
\chi^{-1} ({\bs{k}},\omega) = - i\omega T M({\bs{k}},\omega)/k^2,
\label{chi_M}
\end{equation}
where 
\begin{equation}
M({\bs{k}},\omega) = \left|
\begin{array}{cc}
	0 & {\bs M}_N \\
	\breve{\bs M}_N & \M_{NN}
\end{array}
\right|^{-1} |\M_{NN}|,
\label{M}
\end{equation}
 ${\bs M}_N$ is the row of elements $\{M_1\ldots M_N\}$, 
$\breve{\bs M}_N$ is the column of elements $\{\breve{M}_1\ldots \breve{M}_N\}$,
and  $\M_{NN}$ is the matrix of elements $\{M_{ij}\}, \, i,j=1\ldots N$,
where 
\begin{equation}
\begin{array}{lcl}
	M_n(\bs{k},\omega)& =& \Bigl({\bs{B}}_n(\bs{k},\omega) ; \bs{J}_{\bs{k}}\Bigr),\\[1.2ex]
\breve{M}_n(\bs{k},\omega) &= & \Bigl({\bs{J}}_{\bs{k}}; {\bs{B}}_n(\bs{k},\omega) \Bigr) ,\\[1.2ex]
M_{mn}(\bs{k},\omega) &= &
\Bigl( {\bs{B}}_m;\left[\dot{\bs{B}}_n+i\omega {\bs{B}}_n\right]\Bigr)+
\\[2ex]
& & \hspace*{-16ex}
\left<
\left[\dot{\bs{B}}_m - \dst\frac{\langle\dot{\bs{B}}_m; \bs{J}_{\bs{k}} \rangle_{\omega+i\delta}}
{\langle {{\bs{B}}_m}; \bs{J}_{\bs{k}} \rangle_{\omega+i\delta} } \bs{J}_{\bs{k} }\right];
\left[\dot{\bs{B}}_n+i\omega {\bs{B}}_n\right]
\right>_{\omega+i\delta}\lefteqn{.}
\end{array}
\label{Mn}
\end{equation}

Here ${\bs{J}}_{\bs{k}}=e/m\sum_{\bs{p}}\hbar {\bs{p}}\, n_{\bs{p},\bs{k}}$ is the operator of the   current  density, $n_{\bs{p},\bs{k}}= a^+_{\bs{p}-\bs{k}/2} a_{\bs{p}+\bs{k}/2}$ is the Wigner form of the single-particle density matrix in the momentum representation, 
$\{{\bs{B}}_n\},\,\,n=1\ldots N$, is the chosen set of observables in the form of moments of the density matrix
\begin{equation}
{\bs{B}}_n({\bs{k}})= {\bs{P}}_{{\bs{k}},n} = \sum_{\bs{p}}\hbar {\bs{p}} (E_p/T)^{(n-1)/2} n_{\bs{p},\bs{k}},
\label{Bn}
\end{equation} 
where  ${\bs{J}}_{\bs{k}}=e {\bs{P}}_{{\bs{k}},0} /m$.
In\rf{Mn}, the expressions like
\begin{equation}
(\hat{A};\hat{B}) = \int\limits_0^\beta d\tau \Tr \left\{\hat{A} (-i\hbar\tau) \hat{B}^+ \rho_0\right\}
\label{AB}
\end{equation}
denote Kubo scalar products of operators $\hat{A}$ and $\hat{B}$, where the operator $\hat{A}$ is taken in Heisenberg representation $\hat{A}(t) = e^{i\hat{H} t/\hbar}\hat{A}e^{-i\hat{H} t}$, where 
$\rho_0 = \Z^{-1}\exp[-(\hat{H}-\mu \hat N)/T]$ is the equilibrium statistical operator of the grand canonical ensemble with  $\Z=\Tr\{e^{-(\hat{H}-\mu \hat N)/T}\}$, 
 $\hat N$ is the electron particle number operator. 
The equilibrium correlation function  
\begin{equation}
\left<\hat{A};\hat{B}\right>_z  = \int\limits_0^\infty d t e^{izt} \left(\hat{A}(t);  \hat{B}\right)
\label{laplAB}
\end{equation}
denotes the Laplace transform of the Kubo scalar product of the operators.  

From\rf{epsPichi},\rf{Pi} and\rf{chi_M} one can express the local field factor in terms of correlation functions\rf{Mn} 
containing the observables $B_n$  as 
\begin{eqnarray} \label{G_M} 
G({\bs{k}},\omega)& =&1 +\frac{1}{v(k)}\left[\frac{1}{\chi({\bs{k}},\omega)} - \frac{1}{\chi_0({\bs{k}},\omega)}\right]\\ \nonumber
& =& -i\omega T M({\bs{k}},\omega) +1 - [v(k)\chi_0({\bs{k}},\omega)]^{-1},
\end{eqnarray}
where $M(\bs{k},\omega)$ is given by Eq.\rf{M}.

\subsection{Long-wavelength limit and dynamical collision frequency} 

In the following it is assumed, that the mean free path of an electron between successive collisions $V_{\rm th}/\nu_{\rm eff}$ 
and  the path during the laser period $V_{\rm th}/\omega$ are much smaller than the characteristic  length scale  
of the electric field non-uniformity $L_\nabla$. Here $V_{\rm th} = \sqrt{T/m}$ is proportional to the mean thermal velocity, 
$\nu_{\rm eff}$ is the characteristic  collision frequency of electrons, 
see below Sec.~\ref{sec:relaxtime}. $L_\nabla$ can be of the order of the plasma skin depth 
$l_s=c/\omega_{\rm pl}$, which is created on the surface of a solid target if irradiated by short (subpicosecond) laser pulses, 
where $\omega_{\rm pl}^{2}=4\pi e^{2}n/m$  {is the plasma frequency~\footnote{for longer laser pulses the plasma expansion can lead to an increase of $L_\nabla$ up to length of plasma inhomogeneity  or the laser wavelength}.
In this case, one can disregard the spacial dispersion of the plasma and calculate its optical properties within the \textit{long-wavelength limit}, i.e. for  $k\to 0$. 
For warm dense matter, the above inequalities mean that the considered long-wavelength limit for uniform plasmas is valid for electron temperatures below several hundreds eV~\cite{AndreevTVT03}. 

In the long-wavelength limit is
$ 
\lim_{k\to 0} v(k)\Pi_\RPA 
= \omega_{\rm pl}^2/\omega^2\,.
\label{lw_limit}
$ 
Using Eqs.\rf{epsPichi},\rf{Pi}, one finds a Drude-like expression for the  permittivity
\begin{equation}
\lim_{k\to 0} \eps(\bs{k},\omega) =
\eps(\omega) = 1 - \frac{\omega_{\rm pl}^2}{\omega \left[\omega +i \nu(\omega)\right]},
\label{Drude}
\end{equation} 
where the dynamical collision frequency
\begin{eqnarray}
\nu(\omega)& =& -\frac{\omega_{\rm pl}^2}{i\omega} \lim_{k\to 0} G({\bs{k}},\omega) \\
&= &\omega_{\rm pl}^2 T M(0,\omega) + i(\omega - \omega_{\rm pl}^2/\omega)
\label{nu}
\end{eqnarray}
is, generally speaking, a complex quantity which is closely related to the effective collision frequency of electrons (see below).

It should be noted, that in the long-wavelength limit there is no difference between longitudinal and transverse permittivities, and expression\rf{Drude}, originally derived for the longitudinal permittivity, is also valid for the transverse one.

In accordance with Eqs.\rf{nu},\rf{M}, and\rf{Mn}, the dynamical collision frequency $\nu(\omega)$ is expressed in terms 
of correlation functions. The respective expressions can be derived directly from the general expressions\rf{Mn}, 
as it is done in Refs.~\cite{Reinholz00PRE,Reinholz12PRE}.   It is instructive to describe briefly the derivation directly 
from linear response equations, see  App. \ref{app1}.  Subsequently, the permittivity \rf{Drude} can be calculated 
using the dynamical collision frequency
\begin{equation}
\nu(\omega) = \nu_1 (\omega) r_\omega (\omega), 
\label{nurw}
\end{equation}
\begin{equation}
\nu_1(\omega) = \omega_{\rm au} \frac{\Cf_{11}}{\Nf_{11}},\;
r_\omega(\omega) =\frac{\Nf_{11}}{\Cf_{11}} \frac{1+ i\wstar \sum\nolimits_m \Nf_{1m} \F_m}{\sum\nolimits_m \Nf_{1m} \F_m},
\label{rw}
\end{equation}  
(for details and elimination of $\F_m$ see App. \ref{app1})
  with the following designation of dimensionless correlation functions and response parameter
\begin{eqnarray}
&&\Nf_{nm}= 
\frac{( \hat{\bs{P}}_n ;\hat{\bs{P}}_m )}{mnT},\;
\, \Cf_{nm}(\omega)= \frac{\langle {\hat{\dot{\bs{P}}}}_{n} ; 
{\hat{\dot{\bs{P}}}}_{m}\rangle_{\omega+i \delta}}{mnT\omega_{\rm au}},\nonumber \\
&& F_m = \F_m  \frac{e{\bs{E}}}{mT},
\label{dimcorr}
\end{eqnarray}
{where $\omega_{\rm au}$ is the atomic unit of frequency, 
$\hbar \omega_{\rm au}= E_H = me^4/\hbar^2 \approx 27.2$ eV is Hartree energy,} and $\wstar = \omega/\omega_{\rm au}$ is the dimensionless frequency.

In Eq. (\ref{nurw}),  $\nu_1(\omega)$ is the collision frequency  calculated in the one-moment approximation. For this, only one observable $\hat {\bs{B}}_1 = \hat{\bs{P}}_1$ is used in\rf{rhorel}. In order to take into account higher moments of the distribution function (see Eq.\rf{Bn}), the so called renormalization factor $r_\omega(\omega)$ is introduced \cite{Reinholz00PRE,Reinholz12PRE}. 
A low-order expansion of the correlation functions within perturbation theory with respect to the interaction parameter  $ e^2$ may lead to different results if different \textit{reduced} sets of relevant observables are used.   Therefore, partial summation of the perturbation expansion must be performed to obtain correct results for transport coefficients in this case, see~\cite{Roepke88PRA} and references therein. 

Eqs.\rf{nurw} and\rf{rw} determine implicitly an effective collision frequency of electrons  in terms of dimensionless correlation functions $\Cf_{nm}$ through dimensionless response parameters $\F_n$, which are solutions of the system of equations \rf{Fm}. 
In previous works (see for example~\cite{Reinholz00PRE,Reinholz12PRE}) correlation functions with only first and third moments 
of the distribution function\rf{Bn} in the sum\rf{rw} were considered. It was shown, that this leads to an accuracy 
of few \% for the calculation of the renormalization factor, at least at low frequencies ($\omega/\omega_{\rm pl}< 1$) as well as at high frequencies
$\omega >\omega_{\rm pl}$, when $\lim_{\omega\rightarrow \infty} r_\omega(\omega)\to 1$.
  Using these two moments, the solution of\rf{Fm}  allows  to write down a clearly structured expression for $r_\omega(\omega)$ in terms of those correlation functions:
\begin{equation}
\begin{array}{l}
r(\omega) =  \dst\frac{1}{\Cf_{11}}	\frac{1+i\wstar Q_\omega}{Q_\omega},\quad
Q_\omega = \frac{\Af_{33}-2\Nf_{31}\Af_{31}+\Nf_{31}^2\Af_{11}}
{\Af_{11}\Af_{33}-\Af_{31}^2} ,\\[2ex]
\Af_{lm} = \Cf_{lm}-i\wstar \Nf_{lm}, \; l,m \ge 1.
\end{array}
\label{rw3deg}
\end{equation}

For calculation of correlation functions $\Nf_{lm}$, $l,m\ge 1$, the following expressions are used (see Ref.~\cite{Reinholz15PRE}): 
\begin{equation}
\Nf_{lm} = \dst\frac{\Gamma[(l+m+3)/2]}{\Gamma(5/2)}
\frac{I_{(l+m-1)/2}(\emu)}{I_{1/2}(\emu)}, \; l,m\ge 1,
\label{Cf}
\end{equation}
with $\emu=\mu/T$, from which one has $\Nf_{11}=1$ and 
\begin{equation}
\Nf_{31} = \frac{5}{2} \frac{ I_{3/2}(\emu)}{I_{1/2}(\emu)}, \qquad \Nf_{33} = \frac{35}{4} \frac{I_{5/2}(\emu)}{I_{1/2}(\emu)}.
\label{Cf31}
\end{equation}
Here $I_\nu(y)= \Gamma(\nu+1)\int\nolimits_0^\infty \! x^\nu [e^{x-y}+1]^{-1}dx$ are Fermi integrals; the dimensionless chemical potential is expressed via the inverse Fermi integral $X_{1/2}(x)$ reverse to the Fermi integral $I_{1/2}(x)$ as 
\begin{equation}
 \emu=X_{1/2}\left(2\ef^{3/2}/3\right).
\end{equation}
In the  non-degenerate case $I_\nu(\emu)=e^{\emu}$ for all $\nu$ and 
$\Nf_{31} = 5/2, \; \Nf_{33} = 35/4$, see Ref.~\cite{Reinholz12PRE}. 
 
According to the definitions\rf{laplAB} and\rf{dimcorr}, the correlation functions $\Cf_{nm}(\omega)$ are expressed in terms of correlation functions of occupation numbers as 
\begin{eqnarray}
\Cf_{nm}(\omega)&=&\frac{\beta\hbar^2}{mn\omega_{\rm au}} \sum\nolimits_{p,p'}\!
p_z p_z' (\beta E_p)^{\frac{n+1}{2}}(\beta E_{p'})^{\frac{m+1}{2}} \nonumber \\ && \times
\langle \hat{\dot{n_p}};\hat{\dot{n_{p'}}}\rangle_{\omega+i \delta},
\end{eqnarray}
where $\hat n_p=\hat n_{p,k=0}$. Using the time dependence $\hat{\dot{n_p}}=(i/\hbar)[\hat H,\hat n_p]$, the  Hamiltonian\rf{H} 
and the relation $\left<\hat{A};\hat{B}\right>_z  =
\frac{i}{\beta}\int\nolimits_{-\infty}^{\infty} \frac{d\omega'}{\pi}
\frac{\Im{\mathcal{G}_{AB^\dag}} (\omega'+i\delta)}{\omega'(z-\omega')}, 
$
where $\mathcal{G}_{AB^\dag}$ is the thermodynamic Green function, one can express correlation functions $\Cf_{nm}$ in terms 
of four-particle Green functions, containing products of  potentials for electron-ion $V_{\rm ei}(q)V_{\rm ei}(q')$ or electron-electron $V_{\rm ee}(q)V_{\rm ee}(q')$ interactions, see, for example,~\cite{Reinholz00PRE}. 

These Green functions can be evaluated by diagram technique. At the lowest order of interaction the four-particle Green functions are expressed as product of single-particle Green functions and the Born approximation follows. Summation of ring diagrams leads to account for dynamical screening of interaction potential and permits one to avoid artificial cut-offs as adopted in classical KT. Summation of ladder diagrams permits one to account for strong collisions with large-angle scattering, see Ref.~\cite{Reinholz00PRE} for details. It should be noted, that for simplicity it's reasonable to calculate the renormalization factor\rf{rw} in the  Born or screened Born (see below) approximation and take into account strong-coupling effects only in the calculation  of the collision frequency $\nu_1(\omega)$ in a one-moment approximation while calculating correlation function $\Cf_{11}(\omega)$. 

The account of screening of the interaction potential is necessary to avoid divergencies at low frequencies~\cite{Roepke88PRA} 
and to get numerically accurate results at finite frequencies. Account of the dynamical screening via summation of 
ring diagrams~\cite{Reinholz00PRE,ReinholzADP} gives rise to the Lennard-Balescu result for the dynamical collision frequency
 $\nu_1(\omega)$.   Adopting the  dimensionless units 
\begin{equation}
	\label{units}
\tilde{k}=k/k_{\debroil};\; \tilde{r}=k_{\debroil}r;\; \tilde{\omega}=\hbar \omega/T,\; 
k_{\debroil}^{-1}= \debroil =\hbar/\sqrt{mT}
\end{equation}
(here $r$ is any value having dimension of coordinate), it can be written in the following form: 
\begin{equation}
	\label{nuLB}
	\tnu^{\rm LB} (\w) = \frac{i\nu_0 Z}{\w}\int\nolimits_0^\infty \!\! {\k}^2 d{\k} \left[\epsrpa^{-1}({\k},\w)-
	\epsrpa({\k},0)^{-1}\right],
\end{equation}
with $\nu_0=2\sqrt{\hbar \omega_{\rm au}/T}/(3\pi)$ and  $\epsrpa = \epsrpa' + i\epsrpa''$ is the RPA permittivity,
\begin{eqnarray}
	\label{RepsRPA}
	\epsrpa'({\k},\w)&=&1+\frac{\nu_0}{{\k}^3}\left[
	g\left(\frac{\w}{{\k}}+\frac{{\k}}{2}\right)-g\left(\frac{\w}{{\k}}-\frac{{\k}}{2}\right)
	\right],
\\ \nonumber
	\label{g}
	g(x)& =& \int_0^{\infty} \frac{\xi d\xi}{\exp(\xi^2/2-\emu)+1}\ln\left|\frac{x+\xi}{x-\xi}\right|;
\end{eqnarray}
\begin{equation}
	\label{IepsRPA}
	\epsrpa''({\k},\w)=\frac{\nu_0}{{\k}^3}
	\ln\left[
\frac{1+\exp[\emu-1/2\left(\w/{\k}-{\k}/2\right)^2]}{1+\exp[\emu-1/2\left(\w/{\k}+{\k}/2\right)^2]}.
	\right].
\end{equation}

These formulas\rf{RepsRPA}-(\ref{IepsRPA}) are for  plasmas at abitrary degeneracy and  were derived in~\cite{Arista84}. 
For non-degenerate plasmas ($\ef  \ll 1$) they go into the form
\begin{multline}
	\epsrpa'({\k},\w)= 
		1 + \sqrt{2} \,\frac{\w_{\rm pl}^2}{{\k}^3}\left[
	D\left\{\frac{1}{\sqrt{2}}\left(\frac{\w}{{\k}}+\frac{{\k}}{2}\right)^2\right\}
	\right. \\[1.2ex] \left.
	-D\left\{\frac{1}{\sqrt{2}}\left(\frac{\w}{{\k}}+\frac{{\k}}{2}\right)^2\right\}\right],
	\label{ReepsRPAgas}	
\end{multline}
\begin{equation}
	\epsrpa''({\k},\w)= \sqrt{2} \,\frac{\w_{\rm pl}^2}{{\k}^3}\sinh\left(\frac{\w}{2}\right)\exp\left[
	-\frac{\w^2/\k^2}{2} -\frac{{\k}^2}{8}\right],
		\label{IepsRPAgas}
\end{equation}
where $D(x) = e^{-x^2}\int\nolimits_0^x\! e^{t^2} dt$ is the Dawson integral.

The Drude-like expression\rf{Drude} with the dynamical collision frequency\rf{nuLB}  describes the permittivity in the whole frequency range. Particularly, it gives the plasmon peak near the plasma frequency $\omega_{\rm pl}$. Results very close to that obtained by
Eq.\rf{nuLB}, but with a lack of details for $\eps(\omega)$
near $\omega=\omega_{\rm pl}$, can be obtained within a  simpler approach by using a statically screened Coulomb potential (Debye potential)~\cite{Roepke88PRA} in Born approximation. Thus one  obtains:
\begin{equation}
	\label{nuBsc}
	\tnu_{\rm D} (\w) = -\frac{i\nu_0 Z}{\w}\int\limits_0^\infty 
	\,\frac{{\k}^2 \!\left[\epsrpa({\k},\w)-
	\epsrpa({\k},0)\right]  d{\k}}{[1 + \k_{\rm D}^2/{\k}^2]^2},
\end{equation}
where $\k_{\rm D}$ is the screening length, as obtained for arbitrary degeneracy~\cite{ReinholzADP},
\begin{equation}
\k_{\rm D}^2 = 3/4 [\R_{\rm D}^2 \ef^{3/2}]^{-1} I_{-1/2}(\emu),
\label{kappascreen}
\end{equation}
and $\R_{\rm D}=R_{\rm D}/\debroil$ is dimensionless Debye radius, $R_{\rm D}=V_{\rm th}/\omega_{\rm pl}$.

In order to put expression\rf{nuBsc} in more explicit form to get similar expressions for higher order correlation functions we rewrite it, following~\cite{Reinholz12PRE}, as: 
\begin{multline}\label{nuBsc2}
\tnu_{\rm D}(\omega)=i \omega_{\rm au}Z/(3\pi^2) \\  \times \int_0^\infty \!dy \! f_{\rm scr}(y) \!
\int_{-\infty}^\infty \frac{dx}{x} 
\frac{1}{w+i \delta -x } 
\ln \left[\frac{1+e^{\emu-(x/y-y)^2}}{1+e^{\emu-(x/y+y)^2}}\right],
\end{multline}
where $w=\w/4$, $y={\k}/(2\sqrt{2})$ and $f_{\rm scr}(y)$ is screening function, which for the case\rf{nuBsc} of statical screening can be written as
\begin{equation}
f_{\rm scr}(y) = y^3/[y^2 + \k_D^2/8].
\label{fscr}
\end{equation}

Using the Sokhotski-Plemej formula
$[w-x+i\delta]^{-1} = \frac{\mathcal{P}}{w-x } -i\pi \Res(x=w)$, the expressions for real and imaginary parts of 
$\tnu_{\rm D}(\omega)=\tnu_{\rm D}'(\omega)+i\tnu_{\rm D}''(\omega)$ can be written as 

\begin{multline}\label{nuBsc2Re}
\tnu_{\rm D}'(\omega)=\dst\frac{\omega_{\rm au}Z}{3\pi w} \int\limits_0^\infty \!dy  f_{\rm scr}(y) \!   
\ln \left[\frac{1+e^{\emu -(w/y-y)^2}}{1+e^{\emu-(w/y+y)^2}}\right],
\end{multline}

\begin{equation}\label{nuBsc2Im}
\tnu_{\rm D}''(\omega)= \dst\frac{ \omega_{\rm au}Z}{3\pi^2 w}\int\limits_0^\infty  \!dy  f_{\rm scr}(y) \!  
\left[\sum\limits_{\delta=\pm 1} \I_{11}^\delta(y)-
2\I_{11}^0(y)\right], 
\end{equation}
\begin{equation}
\I_{11}^l=
\int\limits_0^\infty \dst\frac{d\xi}{\xi}\sum\limits_{\sigma=\pm 1}\sigma
\ln\left[
1+e^{\emu -[\xi+\sigma (y + lw/y)]^2}\right],
\label{I11}
\end{equation}
$l=0,\pm 1$.

According to\rf{rw}, the above expressions for $\nu_1(\omega)$ are equal to the correlation function $\Cf_{11}(\omega)$ 
multiplyed by $\omega_{\rm au}$. The expressions for correlation functions $\Cf_{nm}(\omega)$ for $n$ or (and) $m \, > 1$ 
in screened Born (or Debye) approximation have a form similar to Eq.\rf{nuBsc2} (see Ref.~\cite{Reinholz12PRE}) and are given 
in App. \ref{app2}.

\subsection{Effective screening parameter}

In this section we show how the screening length for statical screening can be derived from expression\rf{nuLB} for the case of dynamical screening. In addition, this approach gives us the possibility to introduce dynamical screening into higher order correlation functions\rf{Cnm2ei},\rf{Cnmi}.
Using a formal comparison with\rf{nuLB}, we get the following expression for the screening function in the case of dynamical screening:
\begin{equation}
f_{\rm scr}(y,w)= \epsrpa^*(y,w)/[y \epsrpa'(y,0)|\epsrpa(y,w)|^2],
\label{scr_dyn}
\end{equation}
where $\epsrpa$ is the RPA permittivity\rf{RepsRPA},\rf{IepsRPA}, which in new variables can be rewritten in an equivalent form as
\begin{multline}
\epsrpa(y,w) = 1 + \dst\frac{\sqrt{\w_{\rm au}}}{8\sqrt{2}\pi}\frac{1}{y^3}
\left[-\sum\limits_{\delta=\pm 1} \I_{11}^{\delta} (y) \right. 
\\[2ex]
\left.
+ i\pi\ln\left(\frac{1+\exp[\emu-(w/y-y)^2]}{1+\exp[\emu-(w/y+y)^2]}\right)
\vphantom{\sum\limits_{\delta=\pm 1}}\right],
\label{epsRPA2}
\end{multline}
where $\I_{11}^{\pm 1}$ is given by\rf{I11}.

Taking in mind, that in the considered case of dynamical screening the  screening function $f_{\rm scr}(y,w)$\rf{scr_dyn} is 
a complex function~\footnote{Note, that in the case of statical screening $f_{\rm scr}$ is dependent only on $y$, rather than on $y$ and $w$, see\rf{scr_dyn_HT},\rf{scr_dyn_LT} and discussion below these formulas}, one can rewrite the expressions for real and imaginary parts of the correlation functions {stipulated by electron-ion interactions as
\begin{equation}
\begin{array}{l}
	{\Cf'}_{nm}^{\rm ei}= {\Cf'}_{nm}^{\rm ei}(f_{\rm scr}')-
{\Cf''}_{nm}^{\rm ei}(f_{\rm scr}''),\\[1ex]
	{\Cf''}_{nm}^{\rm ei}= {\Cf''}_{nm,\rm D}^{\rm ei}(f_{\rm scr}')+
{\Cf'}_{nm}^{\rm ei}(f_{\rm scr}''),
\end{array}	
\label{Cnmdyn}
\end{equation}
where designations ${\Cf'}_{nm}^{\rm ei}(f_{\rm scr}')$ and ${\Cf''}_{nm}^{\rm ei}(f_{\rm scr}')$ (where supperscripts ``{\rm ei}`` 
designate e-i interactions) means that in the respective expressions\rf{Cnmr} and\rf{Cnmi} for real and imaginary parts of correlation functions the real part of screening function\rf{scr_dyn} is substituted for $f_{\rm scr}$, and similarly designations ${\Cf'}_{nm}^{\rm ei}(f_{\rm scr}'')$ and ${\Cf''}_{nm}^{\rm ei}(f_{\rm scr}'')$ means substitution of imaginary part of $f_{\rm scr}$\rf{scr_dyn}.


One can show from\rf{scr_dyn} and\rf{epsRPA2}, that in the non-degenerate low-frequency case, i.e. at $\ef\ll 1$ and $w\ll 1$:
\begin{equation}
\dst f_{\rm scr}\approx \frac{y^3}{(y^2+1/(8\R_{\rm D}^2))^2}
\left[
1 - i\frac{\sqrt{\pi}w/y}
{1+8\R_{\rm D}^2 y^2}
\right],
\label{scr_dyn_HT}
\end{equation}
and  in the degenerate low-frequency case, i.e. at $\ef\gg 1$ and $w\ll 1$:
\begin{equation}
\dst f_{\rm scr}\approx \frac{y^3}{(y^2+3/(16 \R_{\rm D}^2 \ef))^2}
\left[
1 - i\frac{3\pi w /(2y\sqrt{\ef})}
{3+16\R_{\rm D}^2 y^2\ef}
\right].
\label{scr_dyn_LT}
\end{equation}

In the low-frequency case the main contribution to the integrals like\rf{Cnmr},\rf{Cnmi} comes from $y\sim \sqrt{w}\ll 1$, hence one can disregard the imaginary parts in Eqs.\rf{scr_dyn_HT}, \rf{scr_dyn_LT} and use the following expression in\rf{fscr} to ensure proper interpolation between\rf{scr_dyn_HT} and\rf{scr_dyn_LT}:
\begin{equation}
\k_{\rm D}^2  \approx \k_{\rm D,deg}^2  = [\R_D^2 (1+2\ef/3)]^{-1},
\label{kappaD}
\end{equation}
which also gives an interpolation of Eq.\rf{kappascreen}. It ensures good agreement between calculations using expressions\rf{nuLB} and\rf{nuBsc}, see Sec.~\ref{s:results} below.  A similar expression
$\k_{\rm D}^2 = \R_{\rm D}^{-2}/[1+\ef^4]^{1/4}$
was introduced in Ref.~\cite{Gericke10PRE}, but it gives  wrong asymptotics at low temperatures and less agreement 
when comparing with results obtained from Eqs.\rf{nuLB} and\rf{nuBsc}.

For strongly coupled plasmas, the perturbative approach, which is the basis for LRT , is, 
generally speaking, not valid. In this case, the screening parameter can be different from that described above. 
In Ref.~\cite{Skupsky87PRA} it was argued, that one should use the maximum of the Debye length and interatomic distance\rf{R0} 
as screening length in dense coupled plasmas. That means, that formula\rf{kappaD} in strongly coupled plasmas could be rewritten as
\begin{equation}
\k_{\rm D}^2 = \dst\min{\k_{\rm D,deg}^2,\k_{\rm max}^2},  
\label{kappaD2}
\end{equation}
\begin{equation}
\k_{\rm max}^2=  8\ef/(18\pi Z)^{2/3}.
\label{kappamax}
\end{equation}
Taking in mind\rf{kappaD2} and\rf{fscr}, one can suppose that in the case of dynamical screening\rf{scr_dyn} the screening function $f_{\rm scr}$ will also be  restricted from below by the value
\begin{equation}
f_{\rm scr,min}=y^3/[y^2 + \k_{\rm max}^2/8],
\label{fscrmin}
\end{equation}
where $\k_{\rm max}$ is given by\rf{kappamax}.

The importance of taking into account the screening of Coulomb potential was underlined in a recent paper~\cite{Faussurier14poP}. 
Unlike Ref.~\cite{Faussurier14poP}, our approach permits one to take into account different versions of screening and 
does not need special ``Drude-like infrared regularization``~\cite{Johnson2006JQSRT,Kuchiev08PRE} at small frequencies of radiation.

It is interesting to note, that  expression\rf{kappaD} can be rewritten as
\begin{equation}
\k_{\rm D,deg}^2 = \dst\frac{8\Gamma_{ei}}{(2\sqrt{3}\pi Z)^{2/3}} \frac{\ef}{1+2\ef/3}.
\label{kappaD3}
\end{equation}
Taking in mind expression\rf{Gamma} for generalized electron-ion coupling parameter $\Gamma_{\rm deg}$ (see Sec.~\ref{sec:nueff} below), one can conclude from expressions\rf{kappamax} and\rf{kappaD3}, that the restrictions of screening occurs at 
$$
\Gamma_{\rm deg} > 1/3.
$$
}

\subsection{High-frequency asymptotics and inverse bremsstrahlung}
\label{s:HF}

For $\w \gg \emu$ and $\w \gg \w_{\rm pl}$ it can be shown, that one can disregard $\k_{\rm D}^2$ in Eq.\rf{nuBsc2Re} 
(the respective terms are exponentially small, $\sim e^{\emu-w}$) and rewrite it as 
\begin{equation}\label{nuBscRe_winf}
\tnu_{\rm D}'(\omega)= \frac{\omega_{\rm au}Z}{3\pi w} \int_0^\infty  \frac{dy}
{y} 
\ln \left[\frac{1+e^{\emu -(w/y-y)^2}}{1+e^{\emu-(w/y+y)^2}}\right].
\end{equation}

With account of the inequality $w\gg 1$, the expression\rf{nuBscRe_winf} can be simplified and written down in terms of asymptotic series with respect to the parameter $w^{-1}$. With account for only leading order terms this can be written 
as~\footnote{It's interesting to note, that the derivation of high-frequency inverse bremsstrahlung on the base of consideration of emission of an electron scattering at Coulomb center, conducted at~\cite{Krainov01JETP,Krainov00JPB}, gave rise to dependence $\nu_{\rm eff}\sim \omega^{-2/3}$ of effective frequency of collisions on laser frequency instead of dependence $\sim \omega^{-3/2}$ as it follows from our consideration, see\rf{nuBscRe_winfd}. The reason of this discrepancy is not presently clear.}  
\begin{equation}\label{nuBscRe_winfd}
\tnu_{\rm D}'(\omega)= \frac{\w_{\rm au}Z}{3\pi w^{3/2}} \int_{-\infty}^\infty  dt
\ln \left[1+e^{\emu -4t^2}\right]
\end{equation}
for the case of arbitrary degeneracy or 
\begin{equation}\label{nuBscRe_winfdd}
\tnu_{\rm D}'(\omega)\approx \frac{2\w_{\rm au}Z \emu^{3/2}}{9\pi w^{3/2}}  \left[1+   \dst \frac{\pi^2}{8\emu^2}+\frac{7\pi^4}{640\emu^4}\right]
\end{equation}
for the highly-degenerate case $\emu > 1$.

From\rf{nuBscRe_winf} one immediately can obtain the  well-known expression~\cite{Schlessinger79PRA} for non-degenerate plasmas
\begin{equation}\label{nuBscRe_winfnd}
\tnu_{\rm D}'(\omega)= \frac{32\w_{\rm au}Z}{9\pi^{3/2} \w}\ef^{3/2} \sinh\left(\frac{\w}{2}\right)K_0 \left(\frac{\w}{2}\right),
\end{equation}
where $K_0(x)=\int_0^\infty dt \exp[-x \cosh( t)] = \int_0^\infty dy\,\exp[-y^2-x^2/(4y^2)]/y  $ is the modified Bessel function of the second kind. Taking in mind the limit $\w \gg 1$, one can derive from\rf{nuBscRe_winfnd} the following asymptotic :
\begin{equation}\label{nuBscRe_winfnd2}
\tnu_{\rm D}'(\omega \gg 1) = \frac{16\w_{\rm au}Z}{9\pi \w^{3/2}}\ef^{3/2} .
\end{equation}
This expression coincides with the first term of a similar expression for degenerate plasmas, Eq.\rf{nuBscRe_winfdd}, if one takes into account that $w=\w/4$ and $\emu \approx \ef$ for the degenerate case. 

For the imaginary part of the permittivity and large $\w$ one has from Eq.\rf{nuBsc2Im} the expression
\begin{equation}\label{nuBscIm_winfd}
\tnu_{\rm D}''(\omega)= -\frac{8\w_{\rm au}Z}{3\pi^2 \w} 
\int\limits_0^\infty dy \, \! f_{\rm scr}(y) \,
\I_{11}^0(\emu,y),
\end{equation}
where $\I_{11}^0$ is given by\rf{I11}; it is not depending on $\omega$.

The dielectric function $\eps (\omega)=[n(\omega)+i c/(2 \omega) \alpha (\omega)]^{2}$ 
 determines the refraction index $n(\omega)$ as well as the absorption coefficient $\alpha(\omega)$, which is related in thermal equilibrium with emission coefficient $j(\omega)$ by  Kirchhoff's law of radiation $j(\omega)=\alpha(\omega)L_\omega(\omega)$, where
 $L_\omega(\omega)$ is the spectral power density of black body radiation. 
 
In the high-frequency limit, where  $n(\omega) \approx 1  $ and $\omega \gg \nu_{\rm D}' $, one has 
\begin{equation}
\alpha(\omega) = \frac{\omega}{c \,n(\omega)} {\rm Im}\, \eps(\omega) \approx \frac{\omega_{\rm pl}^2}{\omega^2 c} \nu_{\rm D}'(\omega),
\label{alpha_eps}
\end{equation}
so that the inverse bremsstrahlung absorption coefficient is directly related to the real part of the dynamical collision frequency obtained above.
 
 Using Eqs.\rf{alpha_eps} and\rf{nuBscRe_winfnd} in the non-degenerate limit, one can write an expression for $\alpha(\omega)$ in the following form:
\begin{equation}
\begin{array}{lcl}
\dst	\frac{c\hbar \alpha}{T}&=& \dst \frac{32}{9}\sqrt{\frac{\pi}{3}}Z\frac{\omega_{\rm au}\omega_{\rm pl}^2}{\omega^3}\ef^{3/2} 
\left(1-e^{-\w}\right)g_{ff}^{\rm Born}(\omega),\\[2ex]
g_{ff}^{\rm Born}&=&(\sqrt{3}/\pi^2)\exp(\w/2)K_0(\w/2),
\end{array}
\label{alpha_Bscnd}
\end{equation}
where $g_{ff}^{\rm Born}$ is
the free-free Gaunt factor  in Born approximation, see Refs. \cite{Wierling01PoP,ReinholzADP,Fortmann05arXiv,Fortmann2005}. 

The expression\rf{alpha_Bscnd} coincides with the result derived in~\cite{Schlessinger79PRA} and with the well-know Bethe-Heitler expression resulting from QED in second order of interaction \cite{Bekefi66}
for a hydrogen plasma in the non-relativistic limit. 


The well-known Kramers formula for the inverse bremsstrahlung absorption \cite{Kramers23PhMag} results with the Gaunt factor $g^{\rm Kramers}_{ff}(\omega)=1$. The same approximation for the Gaunt factor was used in a recent paper~\cite{Faussurier14poP}.

This one-moment Born approximation can be improved taking into account dynamical screening, 
strong collisions, and higher moments in the statistical operator, as discussed earlier. However, in the high-frequency limit, the dynamical screening is not of relevance. Similarly, the renormalization factor $r_\omega(\omega)\to 1$ for $\omega \gg \omega_{\rm pl}$, see Refs. \cite{ReinholzADP,Reinholz12PRE} and\FF{nu_w300_A} below.

Strong collisions have been considered and lead to the famous Sommerfeld result for the Gaunt factor \cite{Sommerfeld49,Wierling01PoP}.
For dense plasmas, the account of ion correlation has a major effect and can be directly included in the Born approximation \cite{Totsuji85PRA}  via the static structure factor $S_{\rm ii}(\k)$, see section~\ref{s:Sii} below.

The standard treatment of the kinetic equation using a relaxation time ansatz, see Sec.~\ref{sec:relaxtime}, fails to describe inverse bremsstrahlung absorption at high frequencies. The frequently used classical kinetic expression for the dynamical conductivity, or the corresponding expression for the dielectric function, are restricted to the low-frequency region since a static, $p$-dependent (and $\omega$-independent) relaxation time cannot be applied to the high-frequency region. Different approaches using Fermi's golden rule have been used \cite{Schlessinger79PRA,LL_Quant_Electrodynamics} to derive expressions for the emission of radiation.
A common treatment unifying both limiting cases, $\omega \to 0$ and $\omega \to \infty$, is missing in KT if the relaxation time approximation is used.

In contrast,  our approach within LRT covers the entire frequency regime consistently, is applicable to the degenerate case~\cite{Kawakami88PRA} and can also be applied to the relativistic regime~\cite{Holl03PhysicaA}.  An important feature of the LRT is the possibility to include medium effects in dense plasmas such as the Landau-Pomeranchuk-Migdal effect~\cite{Fortmann05arXiv,Fortmann2005}.

\subsection{Low-frequency asymptotics}

For $\w\ll 1$ one has the following asymptotics from expressions\rf{Cnmr}
and\rf{Cnmi}
\begin{multline}\label{Cnmrw0}
{\Cf'}_{nm}^{{\rm e}q}(\w)= \dst \frac{4\alpha_q}{3\pi}
\int_0^\infty f_{\rm scr}^q (y) R_{nm}^{{\rm e}q}(0,y) 
\frac{e^{\emu-y^2} dy}{1+e^{\emu-y^2}},
\end{multline}
\begin{multline}\label{Cnmiw0}
{\Cf''}_{nm}^{{\rm e}q}(\w)= \dst\frac{w\alpha_q}{3\pi^2}\int\limits_0^\infty  \! \frac{ dy}{y^2} \,f_{\rm scr}^q(y)
\int\limits_0^\infty \dst\frac{d\xi}{\xi}\sum\limits_{\sigma=\pm 1}\sigma \times\\
 \times
\frac{\partial^2}{\partial\xi^2}\left\{
R_{nm}^{{\rm e} q}\left(\xi,y\right)\ln\left[
1+e^{\emu -(\xi+\sigma y)^2}\right]\right\},
\end{multline}
where superscripts ``{\rm e}q`` designate e-e or e-i interactions; $w=\w/4$.

From these expressions it is seen, that the real part of the correlation functions is independent of $\omega$, while the imaginary part is vanishing proportional to $\omega$.

Disregarding the small imaginary part ${\Cf''}_{11}^{{\rm ei}}$, one  obtains from\rf{Cnmrw0} and\rf{nurw} the following expression for the dynamical collision frequency $\nu$ by LRT in the considered low-frequency limit:
 \begin{equation}
\nu (\w\ll 1) \approx \nuplo \, r_\omega' \,  \Lambda_{LRT} ,\; 
\nuplo = \dst \frac{4\sqrt{2\pi}}{3} \frac{n_e e^4 Z}{\sqrt{m} T^{3/2}},
 \label{nuw0}
 \end{equation}
where $r_\omega'$ is real part of renormalization factor and
\begin{equation}
\Lambda_{LRT} = \dst \frac{3\sqrt{\pi}/4}{\ef^{3/2}}
\int_0^\infty f_{\rm scr} (y) 
\frac{e^{\emu-y^2} }{1+e^{\emu-y^2}} dy
\label{LLRT}
\end{equation}   
is the Coulomb logarithm. In the non--degenerate case and with expression\rf{kappaD} for $f_{\rm scr}$ one has from\rf{LLRT} the following expression:
\begin{equation}
\Lambda_{LRT} (\ef\ll 1) \approx  
(\ln \varkappa^{-1} - C - 1)/2,
\quad \varkappa =\k_{\rm D}^2/8,
\label{LBH}
\end{equation}
coincident with the Brooks-Herring Coulomb logarithm~\cite{blatt57SSPh}, where 
$\k_{\rm D}^2/8$ is given by\rf{kappaD},\rf{kappaD3} and $C \approx 0.5772$ is Euler constant.

\subsection{Ion correlations and screening}
\label{s:Sii}

We now consider the incorporation of ion correlations explicitly according to Eq.\rf{Vei} via the static structure factor.
For the estimation of the ion structure factor for non-crystalline materials, like liquid metals or dense plasmas, 
the following  analytical model~\cite{Thakor11PhChLiq,Bretonnet88PRB}, derived within a one-component plasma model, can be used:
 
\begin{multline}
S_{\rm ii}(k) =\dst \left[1-\frac{3\Gamma_{\rm ii}}{(k R_0)^4 a_2^2}
\left[\cos(k R_0 a_1)+ 2\cos(k R_0 a_2)\right.\right.
\\
\left.\left.   
-3\sin(k R_0 a_1)/(k R_0 a_1) \right]
\vphantom{\dst\frac{3\Gamma_{ii}}{a_2^2}}
\right]^{-1}
,
\label{Sii}
\end{multline}
$$
\begin{array}{l}
a_1= -0.1455\, 10^{-2} \Gamma_{\rm ii} +a_{1Z}(Z), \\[2ex]
a_{1Z}(Z) =
\begin{cases}
0.96, & Z=1\\
1.0, & Z=2\\
1.08, & Z=3\\
1.15, & Z\ge 4
\end{cases}
,\quad 
a_{2}(Z) =
\begin{cases}
1.45, & Z=1\\
1.80, & Z=2\\
2.10, & Z=3\\
2.25, & Z\ge 4
\end{cases}
\end{array},
$$
$\Gamma_{\rm ii} = (Ze)^2/(R_0 T_{\rm i})$, $R_0$ is interatomic distance, see Eq.\rf{R0} above. 
Since this model doesn't incorporate  properties of specific metals,  it can be used for an estimation of $S_{ii}$ for WDM, including Al plasmas.

Besides ion correlations, it is important to take into account the influence on the permittivity of warm dense matter caused 
by the screening of the Coulomb potential and Pauli blocking due to the interaction with bound and valence electrons near the nucleus~\cite{Gericke10PRE} of complex ions.
This can be done by introducing some pseudo-potential instead of the Coulomb potential of the ion. The most simple form of such a pseudopotential is the empty core potential~\cite{Gericke10PRE,Ashcroft78} in the form 
\begin{equation}
V_{\rm ei}(r) = \begin{cases}
Ze^2/r & \mbox{ for $r > r_{\rm cut}$ }\\
0 & \mbox{ for $r \le r_{\rm cut}$ }
\end{cases}
\label{Veicut}
\end{equation}
where the radius $r_{\rm cut}$ is treated as a free parameter which can be fitted to match experimental data on 
transport and optical properties. Taking in mind the respective expression for the Fourier transform 
of the potential\rf{Veicut}, see Ref.~\cite{Gericke10PRE}, one gets a modified expression for the screening function $f_{\rm scr}(y)$
(remember that $y=\k/(2\sqrt{2})$), which takes into account the difference between the screening function for the pure Coulomb potential and the potential\rf{Veicut} in the expressions for correlation functions containing the interactions between electrons and ions, 
\begin{equation}
f_{\rm scr}^{\rm ei} (y) = f_{\rm scr}(y)\cos^2( 2\sqrt{2} y \tilde{r}_{\rm cut}).
\label{fGericke}
\end{equation}
In the case of complex ions expression\rf{fGericke} is taken for $f_{\rm scr}^{\rm i}$ in expressions\rf{Cnmr} and\rf{Cnmi} for the e-i-correlation functions.


\section{Dielectric function from kinetic theory}
\label{sec:relaxtime}

\subsection{Effective collision frequency}

\label{sec:nueff}

The more simple though less common treatment of plasma permittivity can be done using kinetic equations for the single-particle electron distribution function. 
In Ref.~\cite{Reinholz12PRE} it was demonstrated that quantum kinetic equations can be derived within the scope of quantum statistical theory and hence it is  formally equivalent to the method of quantum statistical operator used above.  

On the other hand, frequently used classical kinetic equations within relaxation time approximation  \cite{LL_PhysKin} are, generally speaking, applicable only for low-frequency perturbations, as long as the electron-ion collisions in relaxation time approximation are independent on frequency. Nevertheless, due to it's simplicity this method is widely used in hydrodynamic codes and also it is convenient for the construction of semi-empirical models based on experimental data~\cite{Veysman08JPB,Agranat07JETPHL,Povarnitsyn12AppSS}. 

A slightly more complex, but still elementary approach is based on an approximate solution of the Fokker-Planck equation as proposed in Ref.~\cite{Nersisyan14PRE}.  This  permits  to take into account not only the contribution of electron-ion collisions to the DF $\eps(\omega)$ (as in the case of relaxation time approximation), but also of electron-electron collisions. In accordance 
with \cite{Nersisyan14PRE}, 
the permittivity of plasmas due to intra-band  
transitions is expressed as 
\begin{equation}
\eps(\omega)=1-(\omega_{\rm pl}/\omega_0)^2 \K_0(\omega),
\label{epsK0}
\end{equation}
\begin{equation}
\K_{0}  =\dst\frac{-2i \chii}{\xi_{\omega} \varepsilon^{3/2}_{\mathrm{F}}}
\int_{0}^{\infty } \! F\left(1;\alphaz;i\betaz \xi^3 \right)
\fF(\xi)[1-\fF(\xi)] \xi^7 d\xi,
\label{epsFermy}
\end{equation}
where the function $\K_{0}$  is expressed in terms of the confluent hypergeometric function $F(a;b;z)$;
$
\chii=[1+5/\Zef]^{-1},\,
\xi_{\omega}=(3\sqrt{\pi }/4)(\nu_{\rm eff}^{\rm nd}/\omega), \,
\xi = v/(\sqrt{2}V_{\mathrm{th}}), \,
\alphaz = (\Zef+8)/3, \,
\betaz = \Zef/(3 \xi_\omega)$;  the Fermi function
$\fF(\xi) = \left[1+\exp(\xi^2-\emu)\right]^{-1}$;
$\Zef=Z/\varkappa $ is an effective charge. The function $\varkappa$ is constructed in such a way~\cite{Nersisyan14PRE} that limits at high and low laser frequencies and for non-degenerate~\cite{Brantov08JETP} as well as for degenerate~\cite{Stygar2002PRE} matter are fulfilled: 
\begin{equation}
\begin{array}{l}
\varkappa (\omega) = \varkappa_0/\left[1 + (C/\xi_\omega)^s\right],\quad
\varkappa_0=  Z\left[\widetilde{\gamma}_\sigma^{-1}(Z)-1\right]/5, \\
\widetilde{\gamma}_\sigma=\gamma_\sigma (Z) + \dst\frac{1-\gamma_\sigma (Z)}{1+0.6\ln\left(1+(20\ef)^{-1}\right)},\; \gamma_\sigma=\frac{a+Z}{b+Z},
\end{array}
\label{kappa}
\end{equation}
where the constants were determined as $a=0.87$, $b=2.2$, $C=s=1$.
The value
\begin{equation}
\nu_{\rm eff}^{\rm ndeg}=\nuplo \,\Lambda
\label{nuefnd}
\end{equation}
is an \textit{effective electron-ion collision frequency for non-degenerate plasma}, expressed in terms of Coulomb logarithm $\Lambda$, which can be determined in a wide range of plasma parameters by respective interpolation formulas~\cite{Esser2003CPP,Skupsky87PRA,Stygar2002PRE,Nersisyan14PRE}, see subsection \ref{sec:Lambda}. 

The above formulas ensure proper well known high- and low-frequency skin effect asymptotics for nondegenerate~\cite{VeysmanPoP06,Brantov08JETP}  and for degenerate Lorentz plasmas~\cite{Lee84PhysFluids,Kirkwood09PRB,Basko97PRE,Selchow00NIMA} (disregarding electron-electron collisions). Therfore the calculation of optical properties  is possible for matter in a wide range of parameters of laser and plasmas with arbitrary ion charge.

 We now analyse further the general expression\rf{epsFermy}. A power series expansion of $F$ with respect to its third argument for the case $\betaz \xi^3 \ll 1 $ reads
\begin{equation}
\dst F\left(1;\alphaz ;i\betaz \xi^3\right) = 1 + i\frac{\betaz\xi^3}{\alphaz} -
\frac{\betaz^2\xi^6}{\alphaz(\alphaz+1)}+\ldots,
\label{Fw0}
\end{equation}
and the asymptotic expansion of $F$ in the limit $\Zef \gg 1 $ reads 
\begin{equation}
F\left(1;\alphaz ;i\betaz \xi^3\right) = \dst\frac{1}{1-\betazz} + \sum\limits_{n\geqslant  1}
\frac{1}{\Zef^n}\frac{\betazz P_n(\betazz)}{(1-\betazz)^{2n+1}} ,
\label{FZinf}
\end{equation}
where $\betazz = i\xi^3/\xi_\omega$ and $P_n(\betazz)$ are polynomials of $\betazz$ to the $n$-th power~\cite{Nersisyan14PRE}.

Taking only the first term in the expansion \rf{FZinf}, in the leading order one gets
from \rf{epsFermy} the expression
\begin{equation}
\K_0 (\omega ) =\dst\frac{2}{\varepsilon^{3/2}_{\mathrm{F}}}
\int_{0}^{\infty } \frac{\fF(\xi)[1-\fF(\xi)]}{\xi^3+i\xi_{\omega}} \xi^{7} d\xi,
\label{epsFermyL}
\end{equation}
which coincides with a result, obtained earlier~\cite{Lee84PhysFluids,Kostenko08GSI,Basko97PRE,VeysmanPoP06,Kirkwood09PRB} for the electron conductivity of the Lorentz plasma.

From Eqs.\rf{FZinf},\rf{epsFermy} one gets the expression 
\begin{equation}
\K_0(\omega )= 1 - i\chiii  \xi_{\omega} \varepsilon_{\mathrm{F}}^{-3/2} \left(1+e^{-\emu}\right)^{-1}
\label{eps_winf_F}
\end{equation}
in the high frequency limit $\omega \gg \nu_{\rm eff}$.
From Eqs.\rf{Fw0},\rf{epsFermy} one gets the expression
\begin{equation}
\K_0(\omega )= \frac{3\chir}{\xi^{2}_{\omega}}\frac{I_{7/2}(\emu)}{I_{1/2}(\emu)}  -
\frac{2i\chii}{\xi_{\omega}}\frac{I_{2}(\emu)}{I_{1/2}(\emu)}
\label{eps_w0_F}
\end{equation}
for low frequencies $\omega \ll \nu_{\rm eff}$, where
$\chiii =1+2/\Zef$ and $\chir= (1+5/\Zef)^{-1}(1+8/\Zef)^{-1}$.

From Eq.\rf{eps_winf_F} it follows~\cite{Veysman15elbr}, that in a wide range of degeneracy parameter the effective  electron collision frequency, which determines the dynamical conductivity and the  absorption in kinetic models, can be expressed in the form 
\begin{equation}
\label{nuefdeg}
\nuef = \dst\frac{3\sqrt{\pi} \nuplo}{4 \ef^{3/2}}\frac{1}{1+e^{-\emu}}  \Lambda,
\end{equation}
which reproduces known limiting cases for non-degenerate and highly degenerate plasmas~\cite{Nersisyan14PRE,Veysman15elbr}, in particular, 
Eq.\rf{nuefnd} for a non-degenerate plasmas.

One can rewrite\rf{nuefdeg} in the form
\begin{equation}
\label{nuef2}
\nuef = \sqrt{2/(3\pi)}\omega_{\rm pl} 
\Gamma_{\rm deg}^{3/2} \Lambda(\Gamma_{\rm deg},Z,\varrho),
\end{equation}
where 
\begin{equation}
\label{Gamma}
\Gamma_{\rm deg}= \dst\frac{\Gamma_{\rm ei}}{\ef\left[
4(1+e^{-\emu(\ef)})/(3\sqrt{\pi})
\right]^{2/3}}
\end{equation}
is the generalized electron-ion coupling parameter for plasma at arbitrary degeneracy.  In the non-degenerate case, \rf{Gamma} is the usual expression $\Gamma_{\rm deg} \approx \Gamma_{\rm ei} =Ze^2/(R_0 T)$. In the case of strongly-degenerate ($\ef \gg 1$) plasmas, the respective coupling parameter  depends on the Fermi energy, 
$\Gamma_{\rm deg} \approx (9\pi/16)^{1/3} Ze^2/(R_0 E_F) = 1.21 Ze^2/(R_0 E_F)$.

Alternatively, expression\rf{Gamma}  can be rewritten as
\begin{eqnarray}
\label{Gamma2}
&&\Gamma_{\rm deg}= 2^{-1/6}Z^{2/3}(\hbar\omega_{\rm au}/T)^{1/2}D_\Gamma(\Theta), \\
&&D_\Gamma = \Theta^{1/2}\left[
1+\exp\left[-X_{1/2}\left(2\Theta^{-3/2}/3\right)\right]
\right]^{-1}.
\end{eqnarray}

The function $D_\Gamma(\Theta)$ has a minimum at $\Theta=\Theta_* \approx 0.519$
with $D_\Gamma(\Theta_*)\approx 1.514$. It is slowly varying in the vicinity of $\Theta_*$ ($D_\Gamma \in (1.51;1.72)$ for $\Theta \in (0.27;0.94)$). 
From this fact and Eq.\rf{nuef2} it follows, that in  strongly coupled plasmas the effective collision frequency is proportional to the plasma frequency 
\begin{equation}
\label{nuefmax}
\nuefmax = \C_1\omega_{\rm pl} ,
\end{equation}
with some numerical coefficient $\C_1$, which  is in the order of 1. Effectively, the maximum of the effective collision frequency is a function of the plasma density. The actual value of $\C_1$ can be determined from the comparison with experimental data~\cite{Veysman08JPB,Veysman15elbr,Povarnitsyn12AppSS}.

It should be noted, that if one takes into account not only electron-ion and electron-electron  scattering as in the above consideration, but also electron-phonon interaction and umklapp processes~\cite{Lugovskoy99PRB}, than the effective frequency of collisions and absorption will have some maximum as function of the electron temperature~\cite{Price95PRL,Agranat07JETPHL,Veysman08JPB}.  

\subsection{Coulomb logarithm}
\label{sec:Lambda}

From the considerations given above it is clear, that the differences of various kinetic and semi-empirical models in determining the laser energy absorption are crucially dependent on the determination of the  the Coulomb logarithm $\Lambda$, which is a slowly varying function of density and temperature. Different authors give different expressions for $\Lambda$ within the quantum or the classical kinetic approach. Below some of them will be briefly considered.

A well-known wide-range model for $\Lambda$ was proposed in~\cite{Lee84PhysFluids}. This model was refined by different authors which had proposed expressions that  show better agreement with experimental data. 
A common expression for $\Lambda$ can be written as
\begin{equation}
\Lambda = \CL\ln\left[1 + (\CCL b_{\rm max}/b_{\rm min})^{1/\CL}\right],\label{L}
\end{equation}
where $\CL$ and $\CCL$ are constants, $b_{\rm max}$ and $b_{\rm min}$ are maximum and minimum impact parameters.

An approximation  for $\Lambda$, which  is accurate up to non-logarithmic terms, was suggested in~\cite{Skupsky87PRA} for weakly coupled, high temperature, low density, high $Z$ plasmas (with $b_{\rm max}/b_{\rm min}\gg 1$). 
Extending the interpolation formula proposed in~\cite{Skupsky87PRA} to the case of moderately coupled plasmas, one can rewrite it in the form\rf{L} with  
\begin{equation}
\begin{array}{l}	
b_{\rm max} = \min{ \max{\lambda_{\rm D},R_0}, V_{\rm th}/\omega}, 
\\
b_{\rm min} =  \max{b_{90}(v_1),\lambda_q(v_1)}, 
\end{array}
\label{LSkupsk}
\end{equation}
where $v_{1}=\sqrt{3} V_{\rm th}$, $b_{90}(v)= Ze^2/(mv^2)$ is the impact parameter for $90^o$ scattering, $\lambda_q(v)=\hbar/(2mv)$ is the quantum-mechanical minimum impact parameter. The screening length 
\begin{equation}
\lambda_{\rm D} = R_{\rm D} /\sqrt{1/(1+2\ef/3) +ZT/T_i},
\label{lambdaD}
\end{equation}
 accounts for both the electrons' degeneracy~\footnote{the degeneracy was not accounted for in the original work~\cite{Skupsky87PRA}} via the term $(1+2\ef/3)^{-1}$, see\rf{kappaD2}, and screening by ions via the term  $ZT/T_{\rm i}$.  

The constant $\CL$ in Eq.\rf{L}  is usually taken as $\CL=1/2$, which leads to the result obtained from the classical trajectory approximation~\cite{Skupsky87PRA,Yakubov93UFN,Lee84PhysFluids}.
 Another choice  $\CL=2/3$  ensures proper high-frequency asymptotics (see sec.~\ref{s:HF}) of the real part of the permittivity. 
 The term $V_{\rm th}/\omega$ in\rf{LSkupsk} represents the Dawson-Oberman (DO) correction to the dynamical conductivity in the high-frequency case~\cite{DawsonOberman62PhFluids,DeckerMori94PhPl}. The numerical coefficient of this term was substantiated in~\cite{Skupsky87PRA}.

The constant $\CCL$ in Eq.\rf{L} is determined by the high-temperature asymptotics, where 
$b_{\rm max}/b_{\rm min}\gg 1$ and $\Lambda \approx \ln(b_{\rm max}/b_{\rm min}) + \ln(\CCL)$. 
In~\cite{Skupsky87PRA} the value of $\CCL \approx 0.287$ was proposed. But in~\cite{Skupsky87PRA} only weakly coupled, high temperature plasmas were considered. The consideration of moderately coupled plasmas shows that $\CCL \approx 1$ will be a better choice. More precisely, calculations presented below
have shown, that a better agreement between results of the kinetic approach considered here with the QS ones are obtained with $\CCL = 1$ for frequencies $\omega < \omega_{\rm pl}$ and $\CCL = 0.5$ for frequencies $\omega \gg \omega_{\rm pl}$. We propose the following expression 
\begin{equation}
\CCL \approx 1 - 0.25 \left[\tanh(\omega/\omega_{\rm pl}-5) + 1\right].
\label{CCL}
\end{equation}
which  interpolates between limits $\CCL = 1$ and $\CCL = 0.5$ and is applicable  in the entire frequency range.

The expressions\rf{L},\rf{LSkupsk} have been obtained following the relaxation time approximations of the respective integrals over the velocity space with the electron distribution function. 
More general expressions can be obtained for a velocity-dependent Coulomb logarithm, which can be expressed as~\cite{Adams2007PhPl}
\begin{equation}
\Lambda(v)=\dst\frac 12 \left[
\ln(1+Q) -\frac{Q}{1+Q} - \frac{1}{2}\frac{Q^2}{(1+Q)^2}
\right],
\label{Lp}
\end{equation}
where $Q= (\lambda_{\rm D}/b_{\rm min}(v))^2$ and $b_{\rm min}(v)$ is given by\rf{LSkupsk}, but with the replacement $v_1 \leftrightarrow v$.

The DO-like correction can also be introduced in\rf{Lp} by replacing the above expression for $Q$  by 
\begin{equation}
Q = \min{(\lambda_{\rm D}/b_{\rm min}(v))^2, 8/\w^2} .
\label{Qp}
\end{equation}

The expression\rf{Lp} with only the first and second terms on the right side was derived, for example, in~\cite{Reinholz12PRE} within the first Born approximation using the quantum kinetic equation and the screened Coulomb potential ($b_{\rm min}=\lambda_q(v)$, i.e., the quantum mechanical limit was used in Ref.~\cite{Reinholz12PRE}). The third term in\rf{Lp} arises if one takes into account ionic correlations~\cite{Yakubov93UFN,FortovIakubov06} (in~\cite{Yakubov93UFN} the classical limit $b_{\rm min}=b_{90}$ was used).

The expression\rf{Lp} cannot be extrapolated into the high-frequency region by a simple replacement of $\lambda_{\rm D}$ by $v/\omega$, like in\rf{LSkupsk}. In particular, for small $Q$ (corresponding to large $\omega$) one has from\rf{Lp}: $\Lambda(v) \sim Q^2 \sim \omega^{-4}$ in the case when we consider only the first two terms on the right-hand side, or $\Lambda(v) \sim Q^3 \sim \omega^{-6}$ in the case of three terms.  Note that the correct  asymptote  $\Lambda(v) \sim \omega^{-3/2}$ for $\omega \gg \omega_{\rm pl}$ follows from comparison of\rf{nuefnd} and\rf{nuBscRe_winfd} .

In Ref.~\cite{Stygar2002PRE} an expression for the Coulomb logarithm is derived in second Born approximation. It can be written in the form
\begin{equation}
\Lambda = \dst \ln\left(\frac{b_{mx1}}{\lambda_{q*}}\right)-\frac{1}{2} +
\frac{2b_{90*}}{b_{mx1}}\left[\ln\left(\frac{b_{mx1}}{\lambda_{q*}}\right)-\ln 2^{4/3}
\right],
\label{LStygar}
\end{equation}
where $b_{mx1}=\max{\lambda_{\rm D},R_0}$, $\lambda_{q*}=\hbar/(2m v_*)$, 
$b_{90*}=Ze^2/(mv_*^2)$, $v_*=\sqrt{(7T+2E_F)/m}$~\footnote{Initially, in~\cite{Stygar2002PRE},non-degenerate plasmas are considered, the generalization to degenerate plasmas is formally introduced by means of appropriate expression for $v_*$}.
In accordance with~\cite{Stygar2002PRE}, expression\rf{LStygar} is valid for moderately coupled plasmas, when $\Gamma_{\rm ei} < 1$. 
The extrapolation of expression\rf{LStygar} to the high-frequency case is also difficult because of non-logarithmic terms, which leads to  negative values for $\Lambda$.

Calculations according to  the models discussed above are considered in the following section. 
Before however, it is instructive to compare LRT and KT results for the case of small frequencies $\w \ll 1$, when the dynamical collision frequency $\nu$, calculated by LRT, can be approximated by expression\rf{nuw0}. With the additional requirement $|\nu| \ll \omega $, the function $\K_0$ (see Eq.\rf{epsK0}) calculated by LRT can be written as $\K_0 \approx 1-i\nu'/\omega$ with 
$\nu'$ given by Equ.\rf{nuw0}. In the same domain of parameters, the function $\K_0$ calculated by KT is given, in accordance with\rf{eps_winf_F} and\rf{nuefdeg}, as
$\K_0 \approx 1-i\nu_{\rm eff}/\omega$.
The permittivities calculated by LRT and KT have the same functional form. They differ in the expression for the Coulomb logarithm. The similarity is especially obvious in the non-degenerate case.

Furthermore,  the formal condition for the  applicability of the Born approximation used in the above formulas\rf{Drude},\rf{nurw},\rf{rw},\rf{Cnmr}--\rf{Inm} for LRT  is $mv^2/2 > Ze^2/(\hbar/(mv))$, where $v\approx \sqrt{3T/m}$ is the average electron velocity, or $T> (4/3)Z^2 E_H$, or $2.5 \,\Gamma_{\rm ei}^2\Theta Z^{2/3} <1$. This case corresponds to the condition 
$b_{\rm min} =  \lambda_q(v)$ in expression\rf{LSkupsk} for the minimum impact parameter (``quantum mechanical limit``~\cite{Skupsky87PRA}). As it was shown in~\cite{Skupsky87PRA}, the KT model leading to Eqs.\rf{L}--\rf{lambdaD} can cover the full domain of the parameter $\Gamma_{\rm ei}^2\Theta Z^{2/3}$, i.e. is formally applicable in both classical ($b_{\rm min} =  b_{90}(v)$) and quantum ($b_{\rm min} =  \lambda_q(v)$) limits. On the other hand, as it was shown above, LRT is applicable in a wide frequency range, while KT is well-grounded only for low frequencies $\omega \ll \omega_{\rm pl}$, see Sec.~\ref{sec:results_w} below.

It is important to note, that even at $T< (4/3)Z^2 E_H$ the results for the permittivity from LRT are very close to the ones obtained by KT with semi-empirical expressions for the effective collision frequency if $\omega \ll \omega_{\rm pl}$, i.e. in the frequency range of the applicability of KT, see Sec.~\ref{s:results} below. That means that the LRT constructed above can be used for the extension of the KT to be used in a wider frequency range.


\section{Results of calculations}\label{sec4}
\label{s:results}


 In this section we will present an extensive comparison of numerical results calculated for the LRT and KT approaches.  For the consistency of the LRT calculations, the Kramers-Kronig  relations and sum rules~\cite{RoepkeWierling98PRE} were checked numerically for such conditions as considered in the following subsections. We looked at 
 solid-density aluminum plasmas ($\rho= 2.7$ g cm$^{-3}$) for different temperatures ($T=2,20,300$ eV). 
The accuracy of the s-sum rule was better  than 0.4\% 
for $T=2,20$ eV and 0.6\% for $T=300$ eV.
The  accuracy of the f-sum rule was better than 0.4\% for $T=2,20$ eV and 1\% at $T=300$ eV. 
The Kramers-Kronig relations were checked using  the frequency range $\hbar\omega \in [0.01; 2000]$ eV. The error was found to be  
less than 1.5\%. 
This accuracy can be further improved by increasing the accuracy of the integrals in the respective expressions for the correlation functions.

\subsection{Dependence of the DF on electron temperature}

We investigate first the dependence of the DF on  temperature. Additionally we calculate the absorption coefficient  $A=I_a/I_L$, where   $I_a$ and $I_L$ are  the flux densities   absorbed in matter and incident from vacuum laser, 
respectively~\footnote{Here we consider the case of low-intensity energy fluxes $I_L$, when one can disregard nonlinear effects of dependence of permittivity (or effective collision frequency) on electric field strength~\cite{DeckerMori94PhPl,Kull01PoP,Langdon80PRL}. Besides that, we neclect modifications   of plasma properties by the probe laser pulse. If this is not justified, one should  average the local absorption coefficient over time and space to obtain experimentally measured values $ \bar{A} =\iint I_L(\bs{r},t) A(\bs{r},t)d^3\bs{r}dt /\iint I_L(\bs{r},t) d^3\bs{r}dt  $ }.
The absorption coefficient is related to the DF  via $A=4\Re{\zeta}/|1+\zeta|^2$~\cite{SilinRukhadze13} with $\zeta=1/\sqrt{\varepsilon}$ in the considered long-wavelength limit and at normal incidence of laser radiation. As an example we consider the solid-density aluminum plasma with a constant average ion charge $Z=3$. The plasma  (with $T_{\rm i}=T$) is irradiated by a laser of wavelength $\lambda=0.4$ nm.
The laser  frequency $\omega=4.71\times 10^{15}$ s$^{-1}$ ($\hbar\omega=3.09$ eV) is smaller than the plasma frequency $\omega_{\rm pl}=23.9\times 10^{15}$ s$^{-1}$ ($\hbar\omega_{\rm pl}=15.7$ eV), and therefore $\w \ll 1$ is considered.
The results of our comparative studies are shown in Figs.~\ref{f:eps_T} and~\ref{f:eps_T2}.

\FF{eps_T} shows the real and imaginary part of the permittivity as well as the absorption coefficient in dependence on the  temperature for different approximations. The plasma parameters $\Gamma_{\rm ei}$,  $\Gamma_{\rm deg}$ and $\Theta$ are also given. 
For the frequency considered here, the LRT as described above gives 
practically identical results for the static (screened Born, SB) and dynamical (Lennard-Balescu, LB) screening. The 
The introduction of a minimum screening length in accordance with Eq.\rf{kappaD2} leads to  better agreement of 
the LRT  with  the  semi-empirical model by Povarnitsyn et al.~\cite{Povarnitsyn12AppSS}, which was constructed using data of the reflectivity for
laser heated aluminum in the region of coupled plasmas.

Comparing the two graphs  in SB approximation  demonstrates that it is important to take into account higher moments of the electron distribution function. This can be done using a renormalization factor $r_\omega$, see Ref. \cite{Reinholz12PRE}. The one-moment approximation strongly overestimates the value of imaginary part of $\eps (\omega)$.

Taking into account  strong collisions via the Gould-DeWitt (GDW) model (see~\cite{Reinholz00PRE} for details of T-matrix and GDW calculations) 
and higher moments via the renormalization factor $r_\omega$ 
leads to a good agreement for the absorption coefficient with the  kinetic approximation for $\Gamma_{\rm ei}\le 0.4$. Above temperatures of 100 eV, the effect of strong collisions (GDW) shows an effect of about 20\% in the imaginary part of the permittivity in comparison to LB or statically screened Born approximation. The difference decreases with increasing temperature. 

Using a pseudopotential instead of a screened Coulomb potential as done by Rogers et al.~\cite{Rogers81PRA}, dynamical conductivity and  absorption coefficient are strongly overestimated. It actually leads to an increase of the effective number of conducting electrons  in comparison to the number of valence electrons thus taking into account the influence of core electrons on the permittivity.
 Contrary to that, the model of Gericke {\it et al.}~\cite{Gericke10PRE} leads to an underestimation of the absorption.  


\begin{figure}
\includegraphics[width=0.99\columnwidth]{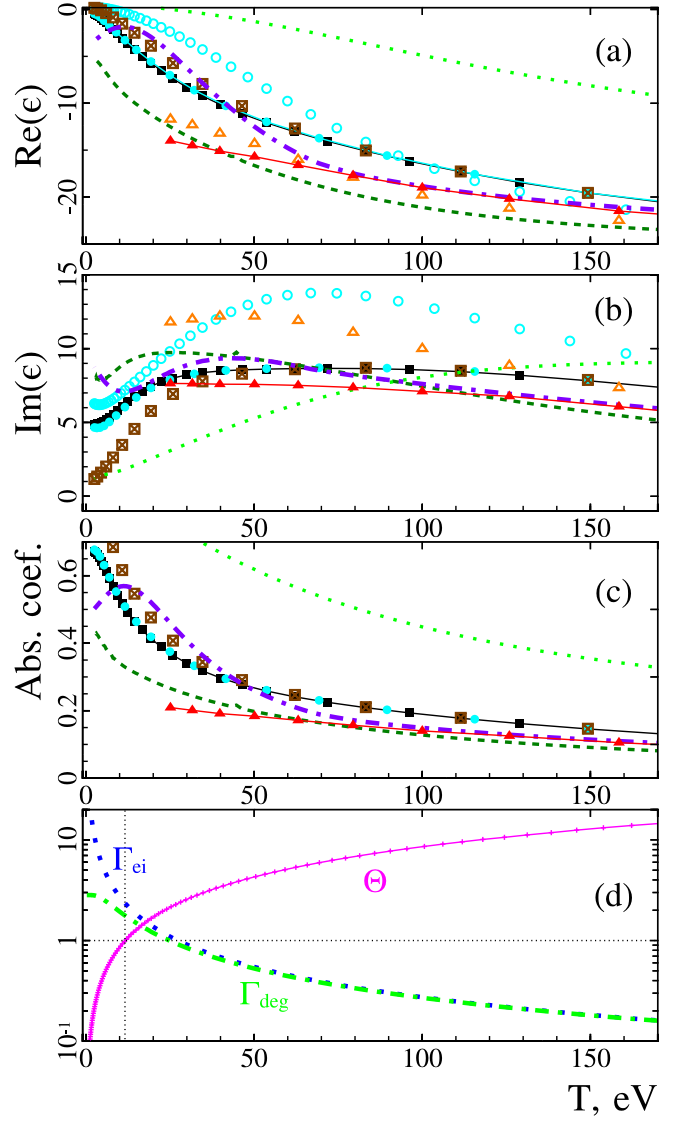}
\vspace{1ex}
\caption{  (Color online).
 Dependence of the real (a) and the imaginary (b) part of $\eps(\omega)$, the absorption coefficient (c) for a frequency of 3.09 eV as well as the coupling parameters  $\Gamma_{\rm ei}$,  $\Gamma_{\rm deg}$ and degeneracy parameter $\Theta$  (d)  on the electron temperature. The vertical dotted line in (d) denotes the Fermi energy.
The following approximations are shown: 
LRT using single-moment SB approximation ($\bs{P}_1$; cyan open circles $\circ$); LRT using  two-moment with SB approximation ($\bs{P}_1, \bs{P}_3$; cyan filled circles $\bullet$); single-moment LRT using SB approximation with restriction of the screening radius from below by $R_0$ (brown  squares $\Box$), LB approximation (black squares $\blacksquare$) and GDW approximation without (open orange triangles $\triangle$) and with (filled red triangles $\blacktriangle$)  renormalization factor $r_\omega$; with account of an empty core pseudopotential as\rf{fGericke} with $r_{\rm cut}=0.4$ \AA  \,\,(dark green dashed line), 
with account for Rogers {\it et al.} pseudopotential~\cite{Rogers81PRA} (green dotted line), and the Povarnitsyn {\it et al.} semi-empirical kinetic model~\cite{Povarnitsyn12AppSS} with the Skupsky~\cite{Skupsky87PRA} Coulomb logarithm (violet dash-dotted lines).}
\label{f:eps_T}
\end{figure}

\FF{eps_T2} shows the  permittivity and the absorption coefficient in a slightly wider temperature range. Different models are compared. For some of the  approaches, the effective collision frequency\rf{nuefdeg} is shown in (d).
The semi-empirical kinetic model of Povarnitsyn {\it et al.}~\cite{Povarnitsyn12AppSS}  (violet dashed-dotted line)
interpolates at $T\sim E_F$ between two different expressions: the phenomenological Drude formula  for metalic plasma ($T < E_F$) and, for $T > E_F$, 
the integral formula for the ideal plasma permittivity of non-degenerate plasmas~\cite{VeysmanPoP06,AndreevTVT03}. 
The two branches of the respective effective collision frequencies are shown (d), the descending curve 
for the non-degenerate plasma ($T > E_F$) and the ascending curve for metallic  plasmas.
Calculations by Cauble and Rozmus~\cite{Cauble95PRE} for a plasma with a step-like density profile are only shown for the absorption coefficient $A$  at normal laser incidence, see   \FF{eps_T2} (c).  In the region of strongly coupled plasmas (at temperatures $T\le 50$ eV), their results considerably underestimate the absorption when comparing with the semi-empirical kinetic or LRT models. 
The  ERR fit formula by Esser et al.~\cite{Esser2003CPP} and the Skupsky~\cite{Skupsky87PRA} model for the Coulomb logarithm $\Lambda$ also underestimate the absorption. The latter is due to a lower  $\Lambda$ and consequently lower $\nu_{\rm eff}$, in comparison with the expressions  given by  Stygar~\cite{Stygar2002PRE} and in the Povarnitsyn~\cite{Povarnitsyn12AppSS} model. 
Note, that the validity of the ERR fit formula, which is based on numerical results of the LRT model and known limiting cases, does not extend to  low temperatures.

On the other hand for $T>10 $ eV, calculations using  Stygar's interpolating expression  for  $\Lambda$ and  the LRT model in two-moments screened Born approximation  are in good agreement with the the semi-empirical model of Povarnitsyn et. al.~\cite{Povarnitsyn12AppSS}. The latter is based on experimental data on the reflectivity of laser-heated aluminum. The ERR fit formula, was originally derived for plasma with singly-charged ions. This could be a reason why it underestimates slightly the dynamical conductivity for aluminum with $Z=3$ at higher temperatures. 
The discrepancies at $T<15 $ eV is connected with the fact that  electron-phonon interactions and absorption in 
metal-like plasmas, which lead to the ascending curve of absorption as function of electron temperature and a maximum 
of absorption near $T=15$ eV, are neither considered in the plasma LRT model described above 
or in the model of Nersisyan  {\it et al.} for the permittivity~\cite{Nersisyan14PRE}   with  
Stygar's  {\it et al.}~\cite{Stygar2002PRE} Coulomb logarithm. An approximate way to account for absorption in metal-like plasmas within the scope of the latter model is proposed in Ref.~\cite{Veysman15elbr}.

\begin{figure}[tbh]
\includegraphics[width=1.0\columnwidth]{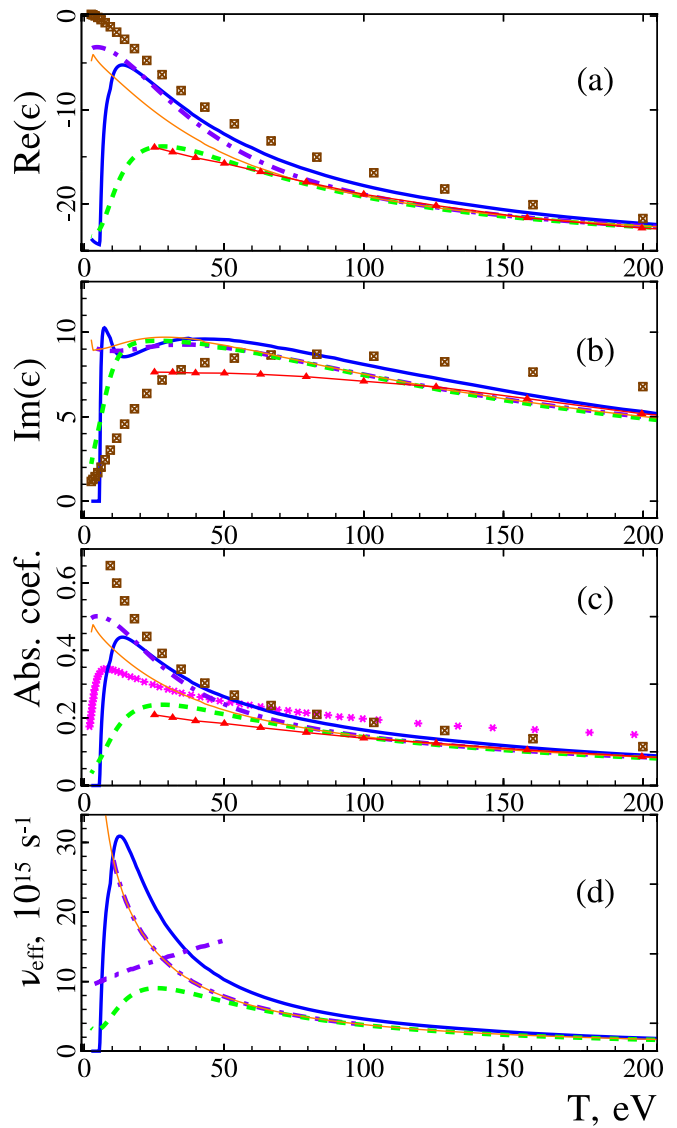}
\vspace{1ex}
\caption{ (Color online).
 The same as on~\FF{eps_T} (a)-(c) and the electron-ion effective collision frequency  $\nu_{\rm eff}$  (d) for an extended temperature range, but not all approximations are shown again (same markers are used) and some additional approximations are added:  
 2-moment SB approximation with the screening model Eq.~\rf{kappaD2} (brown  squares $\Box$), 
 GDW approximation with renormalisation factor $r_\omega$ (filled red triangles $\blacktriangle$), 
 semi-empirical model by Povarnitsyn  {\it et al.}~\cite{Povarnitsyn12AppSS} (violet dashed-dotted line), 
kinetic model\rf{epsK0},\rf{epsFermy}  by Nersisyan  {\it et al.} using different Coulomb logarithms $\Lambda$:   Stygar  {\it et al.} model~\cite{Stygar2002PRE} (solid blue line),  modified Skupsky model~\cite{Skupsky87PRA} 
(thin solid orange line),  ERR fit formula~\cite{Esser2003CPP} (green dashed lines) and  absorption coefficient according to Cauble and Rozmus~\cite{Cauble95PRE} (magenta stars $\star$).}
\label{f:eps_T2}
\end{figure}

\subsection{Frequency dependence of the dynamical collision frequency}
\label{sec:results_w}

The frequency dependence of the complex collision frequency $\nu_{\rm Dr} = \nu_{\rm Dr}' +i \nu_{\rm Dr}''$, defined in accordance with the generalized Drude formula\rf{Drude} by 
\begin{equation}
\begin{array}{lcl}
\nu_{Dr}' (\omega) &=& \omega \Im{n_{\rm e}/n_c/(1-\eps(\omega))},\\
\nu_{Dr}'' (\omega) &=& \omega \Re{1-n_{\rm e}/n_c/(1-\eps(\omega))},
\end{array}
\label{nuDr}
\end{equation}
is shown on\FF{nu_w300_A} and\FF{nu_w300_B}.
Note, that the value of $\nu_{\rm Dr}(\omega)$ defined by\rf{nuDr}  is identical to the value  $\nu(\omega)$ in LRT 
calculations\rf{Drude} and\rf{nurw}, while for the KT calculations with the DF $\eps(\omega)$ 
given by Eqs. \rf{epsK0},\rf{epsFermy},
this complex value $\nu_{\rm Dr}(\omega)$ is quite 
different from the real value of the effective collision frequency $\nuef$, though $\nu_{\rm Dr}'(\omega)$ is 
comparable to $\nuef$, see\FF{nu_w300_A}. 

From\FF{nu_w300_A} it is seen, that for an accurate description of the permittivity by the LRT approach, for frequencies 
lower than the plasma frequency, one needs to take into account not only the first, but also higher moments of the 
electron distribution function when calculating the correlation functions. For $\omega >\omega_{\rm pl}$ the first moment approach is sufficient, i.e., $r_\omega \to 1$ for $\omega \gg\omega_{\rm pl}$. 

It is also seen from\FF{nu_w300_A} that at $\omega > 0.03\,\,\omega_{\rm pl}$ and especially at 
$\omega >0.5\,\, \omega_{\rm pl}$ the imaginary part of $\nu_{\rm Dr}(\omega)$ is not well described by 
the kinetic model. This is due to the fact that the collisional term  is independent on the frequency $\omega$. Contrary, the LRT model gives a consistent imaginary part of $\nu_{\rm Dr}(\omega)$ which satisfies the Kramers-Kronig relations, as discussed already above.

The following\FF{nu_w300_B} demonstrates a comparison of the LRT and different kinetic models using different expressions 
for the Coulomb logarithm, however taking the same parameters as in\FF{nu_w300_A}. For the temperature $T=T_{\rm i}=300$ eV 
the plasma is non-degenerate ($\Theta=26$) and weakly-coupled ($\Gamma_{\rm ei}=0.091$). For the average ion 
charge $Z=3$, the influence of electron-electron collisions is not significant. Therefore the results using the 
expression for $\eps(\omega)$ according to the Povarnitsyn {\it et al.} semi-empirical kinetic 
model~\cite{Povarnitsyn12AppSS} and to the  Nersisyan {\it et al.} model~\cite{Nersisyan14PRE} for plasmas 
with arbitrary $Z$ and similar models for Lorentz 
plasmas~\cite{Basko97PRE} are very close, differences arise from different forms of the Coulomb logarithm. 

\clearpage

\begin{figure}
\includegraphics[width=1.02\columnwidth]{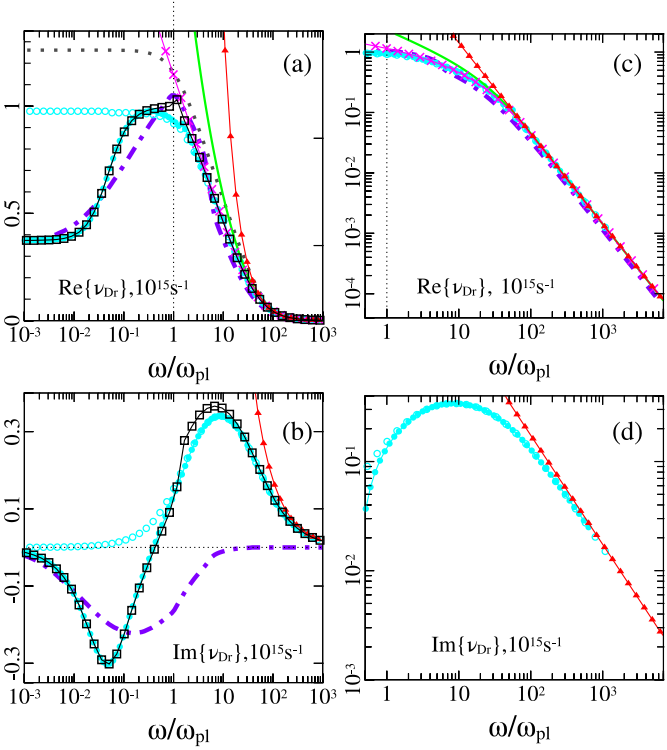}
\caption{(Color online).
 Real (a,c) and imaginary (b,d) parts of the generalized Drude-like collision frequency 
$\nu_{\rm Dr}(\omega)$\rf{nuDr} 
as function of the laser frequency radiating solid-density aluminum plasma at 
$T_{\rm i}=T=300$ eV considered for different frequency ranges. 
Results are shown for the LRT model within 1-moment (cyan open circles $\circ$) and 2-moment (cyan filled circles $\bullet$) SB approximation, the 2-moment LB approximation (black open squares $\Box$), 
the Povarnitsyn  {\it et al.} semi-empirical kinetic model~\cite{Povarnitsyn12AppSS} with the Coulomb 
logarithm\rf{L}--\rf{CCL} and  $\CL=2/3$, without account for screening by ions in\rf{lambdaD} (violet dashed-dotted line),
  $\nuef$ calculated by\rf{nuefdeg} with the same Coulomb logarithm (black dotted line).
 Calculations by asymptotic formulas are shown by thin lines:   $\nu_{\rm Dr}'(\omega)$ from Equ.\rf{nuBscRe_winfnd2} (green line without markers),  Equ.\rf{nuBscRe_winfnd} (magenta line with cross) and Equ.\rf{nuBscRe_winfd} (red line with triangles),  $\nu_{\rm Dr}''(\omega)$  from  Equ.\rf{nuBscIm_winfd}  (red lines with triangles). }
\label{f:nu_w300_A}
\end{figure}

It is seen from \FF{nu_w300_B}, that for  plasma parameters considered here, the second Born approximation for $\Lambda$ as used by Stygar {\it et al.}, Eq.\rf{LStygar} and Ref.~\cite{Stygar2002PRE},  almost coincides with the Skupsky-like 
model\rf{L}--\rf{CCL} for moderate frequencies $\omega<\omega_{\rm pl}$. Furthermore, the expression\rf{Lp} 
without the contribution of ion correlations (i.e. with only two terms on the right hand side of Eq.\rf{Lp}) gives also 
very similar results for small laser frequencies $\omega<0.1\,\,\omega_{\rm pl}$. 

The model \rf{L}--\rf{CCL} is in good agreement with the LRT  for 
$\nu_{\rm Dr}'(\omega)$ in the entire frequency range. 
The account of screening by ions slightly decreases the value of $\nu_{\rm Dr}'(\omega)$, 
compared to the Skupsky-like model shown as dashed-dotted 
curve with the diamond markers. Very similar results are obtained by the ERR fit formula for 
$\Lambda$~\cite{Esser2003CPP,Nersisyan14PRE} at moderate laser frequencies.

\begin{figure}[htb]
\includegraphics[width=0.92\columnwidth]{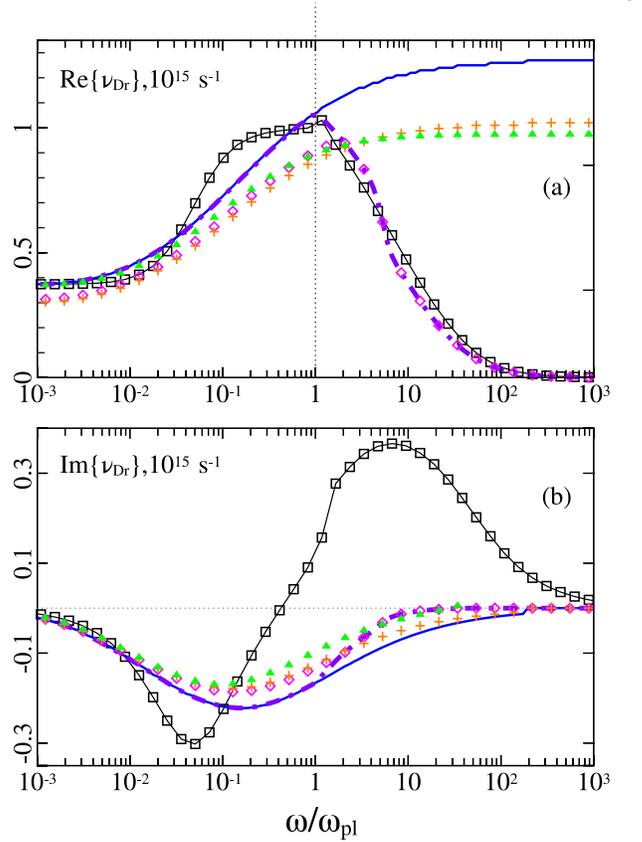}
\caption{(Color online).
 The same as in\FF{nu_w300_A}.
Results are shown for the LRT model within 2-moment LB approximation (black open squares $\Box$), Povarnitsyn  {\it et al.} 
semi-empirical kinetic model~\cite{Povarnitsyn12AppSS} with Coulomb 
logarithm\rf{L}--\rf{CCL} with  $\CL=2/3$, without account for screening by ions in\rf{lambdaD} (violet dashed-dotted line), 
the Nersisyan  {\it et al.} model~\cite{Nersisyan14PRE} with different Coulomb logarithms $\Lambda$: 
the Stygar  {\it et al.} model\rf{LStygar}~\cite{Stygar2002PRE} without account for screening 
by ions in\rf{lambdaD} (solid blue line), the ERR fit formula~\cite{Esser2003CPP} (orange markers +); the model\rf{L}--\rf{CCL} similar 
to Skupsky~\cite{Skupsky87PRA} with  account for  screening by ions in\rf{lambdaD} (red diamonds $\diamond$) and the model\rf{Lp} with only two terms on the right-hand side of\rf{Lp} (filled green triangles $\blacktriangle$).}
\label{f:nu_w300_B}
\end{figure}

\begin{figure}[htb]
\includegraphics[width=0.92\columnwidth]{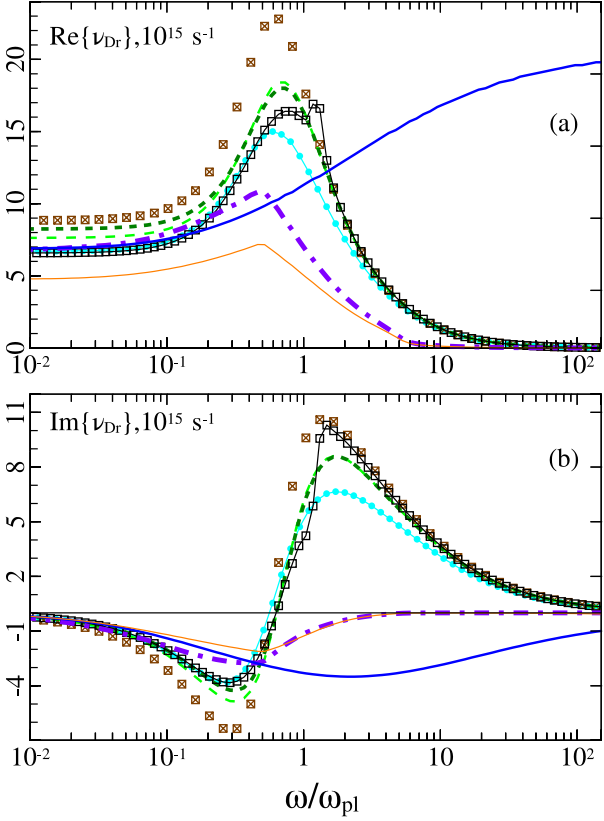}
\caption{(Color online).
 Real and imaginary parts of the generalized Drude-like collision frequency
$\nu_{\rm Dr}(\omega)$\rf{nuDr}, 
as function of the laser frequency radiating solid-density aluminum plasma at  $T_{\rm i}=T=20$ eV. 
Results are shown for the LRT model within the 2-moment LB (black open squares $\Box$) 
and SB (cyan filled circles $\circ$) approximation, 
the same SB approximation with different screening models:  restriction of the 
screening radius from below by $R_0$ 
(brown open squares with cross $\Box$), account for 
ion correlations by the structure factor $S_{\rm ii}$ (green long-dashed curve),  account 
for e-e collisions (dark green short-dashed curve), and 
the Povarnitsyn {\it et al.} 
semi-empirical kinetic model~\cite{Povarnitsyn12AppSS} using the Coulomb 
logarithm\rf{L}--\rf{CCL} with  $\CL=2/3$, (violet dashed-dotted line),  
the Nersisyan {\it et al.} kinetic model~\cite{Nersisyan14PRE} with the
Stygar-like $\Lambda$\rf{LStygar} (solid blue line) and with the Skupsky-like $\Lambda$\rf{LSkupsk} (thin solid orange line).
}
\label{f:nu_w20_A}
\end{figure}


\FF{nu_w20_A} demonstrates the effect of different screening and ion correlations for moderately coupled ($\Gamma_{\rm ei}\sim 1$) solid-density,
partially degenerated aluminum plasmas at temperature $T_{\rm i}=T=20$ eV. For the coupled plasmas
considered here, the restriction of the screening radius from below by $R_0$~\rf{R0} substantially influences the value of $\nu_{\rm Dr}(\omega)$, especially in the region $\omega \in (0.4\, - \,2)\,\,\omega_{\rm pl}$. The real part  as well as $|\nu_{\rm Dr}''(\omega)|$ are considerably increased. The account of ion correlations through $S_{\rm ii}$ also substantially influences the value 
of $\nu_{\rm Dr}(\omega)$, leading to a decreasing peak near $\omega=\omega_{\rm pl}$.
As it was already stated, for the ion charge $Z=3$ considered here, 
the influence of e-e collisions on the dynamical collision frequency is marginal. 

Unlike the case of weakly coupled plasmas\FF{nu_w300_A}, it is seen  from\FF{nu_w20_A}, that for moderately coupled plasmas the agreement between  LRT and kinetic calculations can be observed only for relatively small frequencies $\omega \lesssim 0.3\,\,\omega_{\rm pl}$. For higher frequencies, all kinetic models underestimate the value of $|\nu_{\rm Dr}(\omega)|$.

\FF{nu_w20_B} complements\FF{nu_w20_A}. Additionally it shows calculations with various options for the  Coulomb logarithm\rf{Lp} and the ERR Coulomb logarithm~\cite{Esser2003CPP,Nersisyan14PRE}.
The account of the third term in Eq.\rf{Lp}, responsible for the ion correlations~\cite{Yakubov93UFN}, improves the correspondence with LRT results for 
$\omega \lesssim 0.3\,\,\omega_{\rm pl}$. However, the discrepancies at higher frequencies between both approaches are not removed. Using the ERR fit formula for $\Lambda$ underestimates $|\nu_{\rm Dr}(\omega)|$ in the entire frequency range for the coupled plasmas considered here, which, as it was already stated, can be connected with the fact, that originally this model was formulated for plasmas with singly charged ions.

\begin{figure}[htb]
\includegraphics[width=0.92\columnwidth]{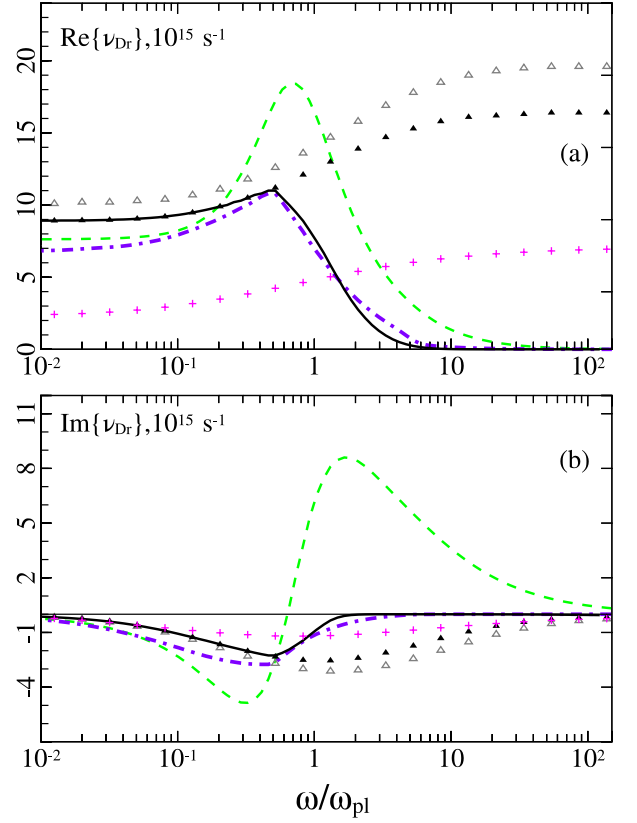}
\caption{(Color online).
 The same as in\FF{nu_w20_A}. Shown are the Nersisyan {\it et al.} kinetic model~\cite{Nersisyan14PRE} with velocity-dependent $\Lambda$, Eq.\rf{Lp},  with two terms on the right-hand side (open gray triangles $\triangle$) or  three terms on the right-hand side (solid black triangles $\blacktriangle$), with account of the DO correction Eq.\rf{Qp} (black line) and with  Coulomb logarithm from 
 ERR fit formula~\cite{Esser2003CPP,Nersisyan14PRE}. Furthermore,e
green dashed and violet dashed-dotted lines have the same meaning as in\FF{nu_w20_A}.
  }
\label{f:nu_w20_B}
\end{figure}

\subsection{Comparison with experimental data}

An interesting experimental quantity is the  dc conductivity $\sigma_{\rm dc}=\sigma (\omega \to 0)$. It is here considered as the dimensionless quantity
\begin{equation}
\dst \sigma^* = \frac{Z\sqrt{m}e^2 }{T^{3/2}}\sigma_{\rm dc} = \frac{3}{4\sqrt{2\pi}} \Lambda_{\rm dc}^{-1}
\label{sigmaDC}
\end{equation}
which is basically proportional to the inverse of the Coulomb logarithm times. 

\FF{sigmaDC_G} shows $\sigma^*$ at a fixed temperature of about 25 keV as function of the coupling parameter $\Gamma_{\rm ei}$.  Several theoretical approaches presented in this paper   are compared with experimental data obtained from rare gas plasmas argon and xenon. Both LRT and  kinetic approaches describe the experimental data points reasonably well, if one takes into account the restriction\rf{nuefmax}  of the maximum of the effective collision frequency. In addition, results are shown taking into account simultaneously  both the ion correlations through $S_{\rm ii}$, the restriction of the screening radius through $R_0$~\rf{R0} and the effect of the interaction of the free, conducting electrons with inner core electrons through a pseudopotential, see Eq.\rf{Veicut} with some radius $r_{\rm cut}=0.5$ \AA. The LRT screened Born calculations, with account of $S_{\rm ii}$ only,  give also satisfying agreement with the experimental data points.

\begin{figure}
\includegraphics[width=0.9\columnwidth]{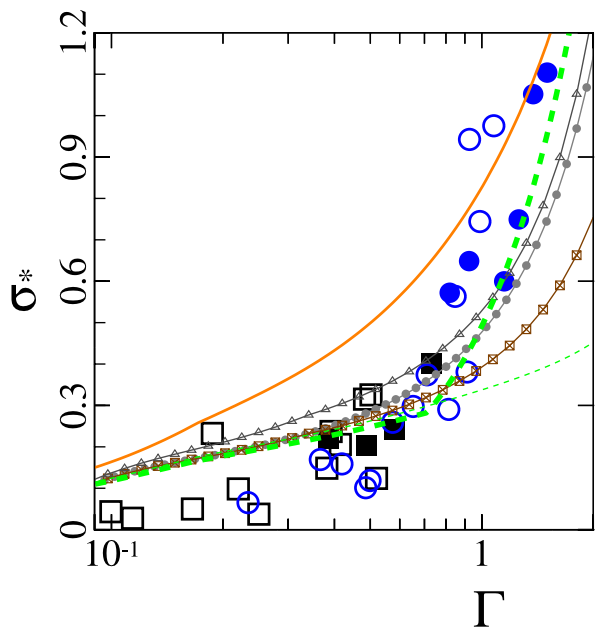}
\caption{(Color online).
 Dimensionless dc conductivity $\sigma^*$, Eq.\rf{sigmaDC},
as function of the coupling parameter $\Gamma_{\rm ei}$ for dense Argon and Xenon plasmas at temperature $T=T_{\rm i} \approx 25$ kK.
Experimental data are shown by large markers: black squares for Argon, blue circles for Xenon, solid markers for experimental points of Ivanov  {\it et al.} \cite{Ivanov76JETP} and open markers for Shilkin  {\it et al.}~\cite{Shilkin03JETP}. Shown are results from LRT model within three-moment screened Born approximation with account for ion correlations (gray line with solid circle), 
with restriction of the screening radius from below by $R_0$ (brown line with  open squares with cross) and with a pseudopotential according to  Gericke  {\it et al.} \rf{Veicut} with $r_{\rm cut}=0.5$\AA (grey line with open triangles).  Furthermore,   calculations from the kinetic model of Nersisyan  {\it et al.} ~\cite{Nersisyan14PRE} is shown using  the Stygar Coulomb logarithm Ref.\rf{LStygar} (thin dashed green line),  applying the restriction of the maximum effective collision frequency by Eq.\rf{nuefmax} with $\C_1=0.9$ (thick dashed line )   and  with the Skupsky-like expression, Eq.\rf{LSkupsk}, for $\Lambda$ (solid orange line).
}
\label{f:sigmaDC_G}
\end{figure}


\FF{Xe_R_ro} shows experimental data on the reflectivity from shock wave fronts in Xe plasmas at different densities. Calculations by the LRT model with and without the restriction of the screening radius from below by $R_0$ are shown. 
It can be seen that results of calculations with account of the restriction of screening in strongly coupled plasmas are closer to the 
experimental points, though still the calculated data are above the experimentally measured reflectivity. 

The plasma density profile of the shock front was assumed to be step-like in the calculations .
Previous studies have shown, that a finite width of the shock wave front~\cite{Reinholz03JPhysA,Reinholz03PRE,Raitza06JPhysA,Winkel09CPP} and 
the contribution of transitions of electrons from bound shells~\cite{Norman15PRE} influence substantially the reflectivity. 
The account of those effects can improve significantly the correspondence of theoretical and experimental results. 
The contribution of bound-bound transitions and the role of plasma inhomogeneities  are investigated in the following section.

\begin{figure}
\includegraphics[width=0.9\columnwidth]{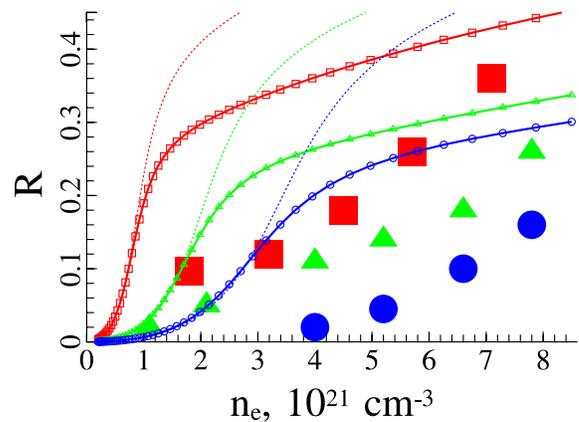}
\caption{(Color online).
 Reflection coefficient $R$ as function of the electrons density $n_{\rm e}$ in reflectivity measurements 
of shock wave fronts in Xenon plasmas, $T=T_{\rm i} \approx 30$ kK.
Experimental data~\cite{Zaporoghets06JPhysA} are shown by large markers: red squares 
for wavelength $\lambda=1.06$ $\mu$m, green triangles for $\lambda=0.694$ $\mu$m, 
blue circles for $\lambda=0.532$ $\mu$m. 
Marked lines (with the same colors and marks for the same $\lambda$) show the LRT two-moment screened Born calculations with account of ion correlations, the restriction of the screening radius from below by $R_0$~\rf{R0}, and with the use of a pseudopotential according to Gericke  {\it et al.} \rf{Veicut} with $r_{\rm cut}=0.5$ \AA\,\, (the same parameters as on\FF{sigmaDC_G}); dotted lines show the same, but without the restriction of the screening radius from below by $R_0$.
}
\label{f:Xe_R_ro}
\end{figure}

\section{Estimates of contributions of bound-bound transitions and plasma inhomogeneities}\label{sec5}

\subsection{Role of plasma inhomogeneities}

Shock waves have a final width $L_S$ of it's front where plasma parameters (for instance temperature, pressure, ion concentration and ionization degree) smoothly change from their upstream (non-perturbed) values  $T_0$, $P_0$, $n_{\rm i,0}$  to their downstream values $T_1$, $P_1$, $n_{\rm i,1}$.  

Estimates on the basis of kinetic equations~\cite{Zeldovich_Raizer} or the Boltzmann H-theorem~\cite{Stupochenko65PMTF} give a value for the  width of the shock wave front of the order of the mean free path of atoms or molecules in the non-perturbed gas. Thus one can write 
\begin{equation}
L_S \approx C_{L_S}/(n_{i,0} \sigma_c),
\label{LS}
\end{equation}
where $C_{L_S}$ is a constant, $\sigma_c$ is the cross section of atomic collisions and 
$n_{i,0}$ is the concentration of heavy particles upstream the flow in the shock wave, 
which is connected with the concentration of heavy particles $n_{\rm i,1}$ downstream the flow by Rankine-Hugoniot relations. It has the following form for a strong shock wave (Mach number $M \gg 1$) in a polytropic gas~\cite{LL_Hydrodynamic,Zeldovich_Raizer}:
\begin{equation}
n_{\rm i,1} = n_{\rm i,0}(\gamma+1)/(\gamma-1), 
\label{Hugonio}
\end{equation} 
where $\gamma$ is the adiabatic coefficient of the gas.

The influence of the plasma inhomogeneity owing to the final width of shock wave front on laser radiation absorption in the plasma depends on the ratio $\kappa_S = L_S/l_s$, where $l_s$ is the plasma skin layer depth. From the solution of wave equation for the uniform plasma with step-like density profile, the permittivity\rf{Drude} and $\omega < \omega_{\rm pl}$, one can write the following expression for $l_s$:

\begin{equation}
l_s \approx \dst \frac{c}{\omega_{\rm pl}}\frac{\sqrt{1+\breve{\nu}^2}}
{\sqrt{1-(\omega/\omega_{\rm pl})^2 (1+\breve{\nu}^2) }} \approx \frac{c}{\omega_{\rm pl}}
\label{ls}
\end{equation}
where $\breve{\nu} = \Re{\nu(\omega)}/\omega$; the second approximate equality is written for the case $\breve{\nu}\ll 1$ and $\omega_{\rm pl} \gg \omega$.

From Eqs.\rf{LS}--\rf{ls} one has the following expression for $\kappa_S$:
\begin{multline}
\kappa_S \approx C_{L_S} \frac{\gamma+1}{\gamma-1}\frac{2\sqrt{\pi Z} e}{\sqrt{m}c\sqrt{n_{\rm i,1}}\sigma_c}\\[2ex] \approx 6\, 10^{-2} C_{L_S} \frac{\gamma+1}{\gamma-1}
\left[\frac{Z}{n_{\rm i,1}/10^{21} \mbox{cm}^{-3}}\right]^{1/2}
\left[\frac{\sigma_c}{10^{-15} \mbox{cm}^{2}}\right]^{-1}.
\label{kappaS}
\end{multline}

Numerical calculations on the basis of the solution of the Navier-Stokes equations give a parameter value $C_{L_S}\approx 4$ for shock waves with Mach numbers $M$ of several units in Argon, see Refs.~\cite{Elizarova05PhysFluids,Jordan11Diser}. 
Similar results for $C_{L_S}$ follow from the numerical solution of the Burnett equations
obtained in Ref.~\cite{Qian2010IJME}.
The simulations on the basis of the solution of the Boltzmann kinetic equations can even give higher values, $C_{L_S}\approx 10$~\cite{Kolobov2012JCP}. 

The value of $\sigma_c$ can be estimated using a fitting formula proposed in Ref.~\cite{Phelps00JPhysB} 
for the total elastic cross section of Argon on Argon atoms with the relative energy $E$:
\begin{equation}
\sigma_c \approx 2.1\left(\frac{E}{ \mbox{eV}}\right)^{-0.4}\left[
1+\left(\frac{E}{15 \mbox{eV}}\right)^2\right]^{0.16}\times 10^{-14} \mbox{cm}^2 ,
\label{sigmafitt}
\end{equation}
which gives the value $\sigma_c \sim 10^{-14}\, \mbox{cm}^2$ for the average relative energy $E \approx 3$ eV. Substituting this value and the values $Z=1$, $C_{L_S}=4$ and $\gamma = 5/3$ into equation\rf{kappaS}, one obtains the following estimate for the experimental conditions depicted on\FF{Xe_R_ro}:
\begin{equation}
\kappa_S \sim 0.1 \left(n_{\rm i,1}/10^{21}\mbox{cm}^{-3} \right)^{-1/2}.
\label{kappaS2}
\end{equation} 

In order to elucidate the influence of the parameter $\kappa_S$ on the absorption of laser energy , let us consider a plasma density profile with a linear ramp:
$n_{\rm i}(x) = 0$ for $x<0$, $n_{\rm i}(x) = \mbox{const} = 
n_{\rm i,1}$ for $x > L_S$,
$n_{\rm i}(x) = n_{\rm i,1} x/L_S$ for $ 0<x < L_S$ (the plasma temperature is assumed to be constant). Under the assumption 
of weak absorption (with $|\nu(\omega)|/\omega \ll 1$) and for overcritical plasma density ($\omega_{\rm pl} \gg \omega$) 
one can express solution of the wave equation for such a plasma profile in terms of the Airy functions $\Ai$ and $\Bi$ 
of the first and the second kind, respectively, and write down the following expression for the value of the change 
of the absorption coefficient (see Ref.~\cite{Veysman01Diser}) $\alpha = A/A_{\rm st}$, 
given by the ratio of real absorption coefficient $A$ for a given plasma profile to the absorption coefficient $A_{\rm st}$ 
for a plasma with step-like density profile $n_{\rm i}(x) = \mbox{const} = n_{\rm i,1}$ for $x > 0$, and $=0$ else:
\begin{equation}
\begin{array}{r}
\dst \alpha =
\frac{\frac{\omega_{\rm pl}^2}{\omega^2}\left[ 1+\frac{2}{\kappa_S}
\int\limits_{-L_0}^{L_1}\!\Phi^2(x)(x+L_0)^2\,d x\right] }{\Phi^2(-L_0)+{\Phi'}^2(-L_0)/L_0}, \\[4ex]
\Phi(x)\equiv \dst C_1 \Ai(x) -C_2 \Bi(x);  \\[1.5ex]
C_1 = \dst \frac{\kappa_S^{1/3}
\Bi(L_1)+\Bi'(L_1)}{\Ai(L_1)\Bi'(L_1)-\Bi(L_1)\Ai'(L_1)},\\[2.5ex]
C_2 =\dst\frac{\kappa_S^{1/3}
\Ai(L_1)+\Ai'(L_1)}{\Ai(L_1)\Bi'(L_1)-\Bi(L_1)\Ai'(L_1)},\\[2.5ex]
L_0 = (\omega/\omega_{\rm pl})^2 \kappa_S^{2/3}, \;
L_1 = [1- (\omega/\omega_{\rm pl})^2]\kappa_S^{2/3},
\end{array}
\label{alpha}
\end{equation}
with $\Phi'(L_0)\equiv \partial \Phi(x)/\partial x|_{x=L_0}$, and similarly for $\Ai'(L_1)$.

From Eq.\rf{alpha} it follows, that in the limit of weak absorption the function $\alpha$ is independent of the absorption mechanism 
(i.e. of $\nu(\omega)$) and depends on only two variables: $L_0$ and $\omega_{\rm pl}/\omega$. 
For $\kappa_S\to 0$ we have  $\alpha\to 1$, and for $\kappa_S\gg 1$ the second term in Eq.\rf{alpha} exceeds the first one, $1$.  
The function $\Phi(x)$ is of the order of $ \Ai(x)$, and the function $\alpha$ is mainly dependent on $L_0$. From that follows, that the lower the ratio $n/n_c = (\omega_{\rm pl}/\omega)^2$ of the maximum plasma density to the critical density $n_c$, the higher is the influence of final width $L_S$ of the plasma density ramp on the absorption. 

The behavior of $\alpha(\kappa_S)$ for different wavelengths corresponding to the experimental ones, see \FF{Xe_R_ro}, 
and different electrons densities are shown in\FF{alpha_KS}. Curves are presented which are calculated using the expression \rf{alpha} 
in the limit of weak absorption as well as curves calculated by the numerical solution of the wave equation for a plasma with 
permittivity given by Eqs.\rf{Drude},\rf{nurw},\rf{rw},\rf{Cnmr},\rf{Cnmi},\rf{Inm} (two-moment screened Born approximation with account of the contribution of electron-electron collisions).  

For the plasma parameters considered here, we have $|\nu(\omega)/\omega|\gtrsim 1$. This is the reason why the 
calculations shown by dashed lines in\FF{alpha_KS}  deviate from the analytical estimates based on Eq.\rf{alpha}. 
In particular, a saturation of the increase of absorption with increasing $\kappa_S$ is seen for $\kappa_S$ 
exceeding some value $\kappa_S=\kappa_{S^*}$. Nevertheless, the above mentioned conclusion about the increase 
of the influence of $\kappa_S$ on absorption, when the value of $n/n_c$ is decreasing, remains true. 
From\FF{alpha_KS} one can see, that the values of $\kappa_{S*}$ is decreasing with the decrease of the  wavelength 
and (or) the diminishment of the plasma density. Particularly, 
$\kappa_{S^*}\approx 1.5,\,\,4,\,\,12$  for $n=1.5, \,\,3,\,\, 6 \times 10^{21}$cm$^{-3}$, respectively, 
and $\lambda=0.53$ \mkm; $\kappa_{S^*}\approx 3,\,\,9$ for $n=1.5,\,\, 3 \times 10^{21}$cm$^{-3}$, respectively, 
and $\lambda=0.69$ \mkm; $\kappa_{S^*}\approx 12$ for $n=1.5\times 10^{21}$cm$^{-3}$ 
and $\lambda=1.06$ \mkm. This could explain, under assumption of final width of shock wave front, why the discrepancy between experimental results and theoretical calculations 
at\FF{Xe_R_ro}, performed with the assumption $\kappa_{S}=0$, is larger for lower plasma densities and for shorter wavelength, 
and why the dependencies of the experimental reflection coefficient $R=1-A$ on the electron density are concave curves, 
while the theoretical calculations under the assumption $\kappa_{S}=0$ lead to convex ones. 

One can see from\FF{alpha_KS}, that the pronounced effect of the nonzero $\kappa_{S}$ on the increase of plasma absorption 
can be achieved only for values of  $\kappa_{S}$ at least of several units, at $\kappa_{S}\gtrsim 3$, 
while theoretical estimates of the width of the shock wave front\rf{kappaS2} gives about 30 times lower values 
$\kappa_{S}\lesssim 0.1$ for $n_{\rm i} \ge 10^{21}$ cm$^{-3}$. 

From the above consideration follows that an increased absorption or a decreased reflection of laser radiation from the 
shock wave front, in comparison with theoretical predictions for $L_s=0$, could be a signature of a considerable 
broadening of the width of the shock wave front due to ionization or excitation processes~\cite{Zeldovich_Raizer,LL_Hydrodynamic}. 
Consequently, such increased absorption could serve as a diagnostic tool to analyze non-stationary processes at the shock wave front.

\begin{figure}
\includegraphics[width=0.9\columnwidth]{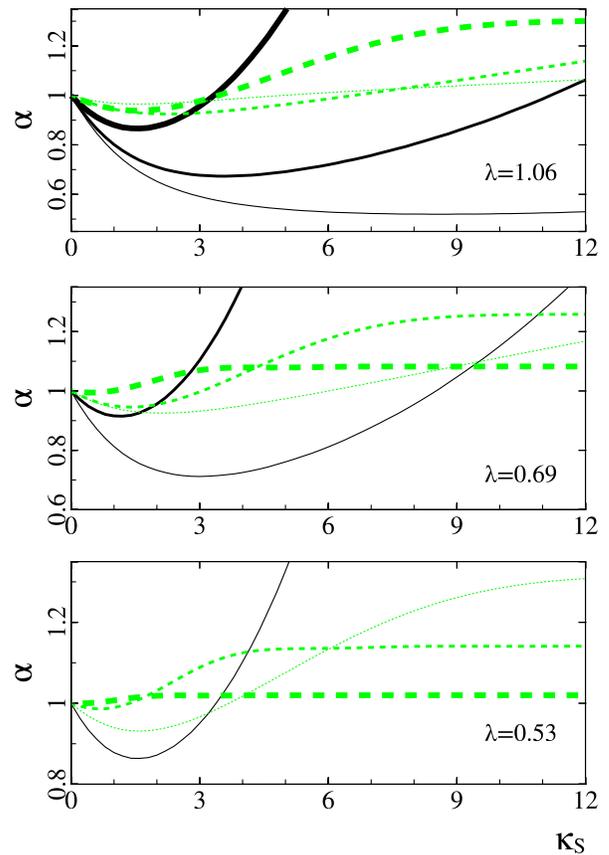}
\caption{(Color online).
 Coefficient $\alpha = A/A_{\rm st}$ as function of the parameter $\kappa_S = L_S/l_s$ of a plasma with linear ramp of it's density profile, for different wavelengths $\lambda=0.53,\,\,0.69\,\,1.06$ \mkm. Different electron densities are considered: 
$n=1.5, \,\,3,\,\, 6 \times 10^{21}$cm$^{-3}$ for thick, medium and thin curves, respectively. 
The calculations in the limit of weak plasma absorption by formula\rf{alpha}  are shown by solid curves. 
Numerical calculations with absorption determined by the three-moment screened Born approximation for Xenon plasmas 
with $T=T_{\rm i} = 30$ kK are shown by dashed curves. (Solid curves corresponding $n/n_c<1$ are absent as long as\rf{alpha} is derived for $n/n_c>1$).
}
\label{f:alpha_KS}
\end{figure}

\subsection{Estimates for the contributions of interband transitions}
\label{s:bb} 

Recently, numerical calculations of the reflectivity of shock compressed Argon and Xenon plasmas utilizing the density functional approach and the Kubo-Greenwood (KG) formula have been performed. They have shown that interband transitions play an essential role and should be accounted for when interpreting the respective experimental results~\cite{Norman15PRE,Lankin15elbr}.
Below a semi-phenomenological estimate of this effect is given. 

One can note, that taking into account  interband transition effects corresponds to the inclusion of  collisions with bound states, in particular atoms. These should be considered  in addition to  electron collisions with free charged carriers,  see~\cite{Kraeft86}. 


With the account for interband transitions, the permittivity can be expressed as
\begin{equation}
\eps(\omega) = \eps_{\rm Dr}(\omega) + \delta\eps_{\rm b}(\omega),
\label{epsfull}
\end{equation}
 where $\eps_{\rm Dr}(\omega)$ is the intraband or Drude-like contribution to the permittivity, given by\rf{Drude}, and  
$\delta\eps_{\rm b}(\omega)$ is the interband contribution to the permittivity. 

Different approaches can be used to determine  $\delta\eps_{\rm b}(\omega)$: besides first-principle calculations using the KG formula, 
see Ref.~\cite{Norman15PRE,Desjarlais02PRE} and their modifications on the basis of the average-atom model~\cite{Johnson2006JQSRT}, 
the Drude-Lorentz (DL) model~\cite{Forouhi86PRB,Benfatto2011PRB,Rakic95AppOpt,Vial2011AppPhysA} 
and it's modification in form of the Critical Points model~\cite{Etchegoin06JChemPhys,Vial2011AppPhysA} 
are widely used for the approximate description of interband contributions to the permittivity.
This way, semi-empirical interpolation formulas for the optical 
properties of a wide class of substances (metals, dielectrics, amorphous materials) are obtained.  

The DL model for $\delta\eps_{\rm b}(\omega)$ can be rewritten in the form
\begin{equation}
\delta\eps_{\rm b}(\omega) = -\omega_{\rm pl,n_i}^2 \sum\limits_{m,n} \dst\frac{F_{mn}^x}{\omega^2-\omega_{mn}^2+i\omega 
\Gamma} f_{\rm F}(E_n)[1-f_{\rm F}(E_m)],
\label{chib}
\end{equation}
where $\Gamma$ is a damping factor, $\omega_{mn}=(E_m-E_n)/\hbar$, 
$\omega_{\rm pl,n_i}^2=4\pi n_{\rm i} e^2/m$. 
$E_m$ is the energy of the $m$-th energy level (or energy of a continuum state in the case of bound-free transitions), 
the matrix element
$F_{mn}^x =  2\,m\,\omega_{mn}|\langle m|x|n\rangle|^2/\hbar$ is the oscillator strength~\cite{Davydov65}, $f_{\rm F}(E) = [1+\exp((E-\mu)/T)]^{-1}$ is Fermi function, the sum in\rf{chib} is over permitted dipole transitions, 
with $l_m = l_n \pm 1$ (where $l_{m,n}$ are the respective orbital quantum numbers).

Unlike the usual DL model, Eq.\rf{chib} contains the factor $f_{F_{mn}}\equiv f_{\rm F}(E_n)[1-f_{\rm F}(E_m)]$ which accounts 
for the population of the energy levels and the Pauli blocking principle. For $\Gamma\to 0$
the imaginary part of Eq.\rf{chib} represents a sum of $\delta$-functions $\delta (\omega+\omega_{mn})+ \delta (\omega-\omega_{mn})$. In this case, 
similarly as it was shown in Ref.~\cite{Moseley78AmJPhys}, 
one can derive from Eq.\rf{chib}  an expression for the imaginary part of $\delta\eps_{\rm b}(\omega)$ which is 
equivalent to the KG formula.  

In the case of laser radiation frequencies much below the transitions frequencies,  $\omega \ll |\omega_{mn}|$ (that is the case of 
reflectivity measurements depicted at\FF{Xe_R_ro}) and for $\Gamma/\omega \ll 1$,  the contribution to the imaginary part of 
$\delta\eps_{\rm b}(\omega)$ is close to $0$. The main influence of the interband transitions on the reflectivity 
(or absorption) comes from 
$\Re{\delta\eps_{\rm b}(\omega)}$. In accordance with Eq.\rf{chib}, the contribution of the $j$-th interband transition to 
$\Re{\delta\eps_{\rm b}(\omega)}$ is proportional to 
\begin{equation}
\delta\eps_{{\rm b},j}'(\omega) = (\omega_{\rm pl,n_i}/\omega_{mn,j})^2 F_{mn,j}^x 
f_{F_{mn,j}}
[1-\omega^2/\omega_{mn,j}^2]^{-1}.
\label{chibj}
\end{equation}

For $\omega \ll |\omega_{mn}|$ the contribution\rf{chibj} depends only weakly on the frequency $\omega$. 
However, it can be dependent on the plasma density $n_{\rm i}$ via the expressions 
$\omega_{\rm pl,n_i}$, $\omega_{mn}$, and $f_{F_{mn}}$. One can assume that $\omega_{mn}$ and $f_{F_{mn}}$ are slowly dependent on $n_{\rm i}$ in comparison with $\omega_{\rm pl,n_i}^2 \sim n_{\rm i}$.
Otherwise, the energy gap $\Delta E_{bf}$ between bound pair excited states (with energies $E_n <0$) and free states of electron gas (with energies $E_f > 0$) increases approximately proportionally to $n_{\rm i}^{1/3}$,
\begin{equation}
\Delta E_{bf} \approx 4.6 \mbox{ eV } Z \left(n_{\rm i}/10^{21}\mbox{cm}^{-3}\right)^{1/3},
\label{Deltabf}
\end{equation}
see Refs.~\cite{Lankin09CTPP,Lankin09JPhysA}. 

Because only states with energies $E_n < -\Delta E_{bf}$ can contribute to interband transitions, 
 the number of transitions contributing to $\delta\eps_{\rm b}(\omega)$  
decreases with increasing of the plasma density, in accordance with\rf{Deltabf}. 
Taking in mind, that transitions from the upper excited levels can give the main contribution to the interband absorption 
(see, e.g., Ref.~\cite{Johnson2006JQSRT}), such a decrease can be rather essential and could compensate the linear increase of $\omega_{\rm pl,n_i}^2$ with $n_{\rm i}$. 

\FF{Xe_R_ro} illustrates, that the simplest assumption of a constant value independent on density, 
$\Re{\delta\eps_{\rm b}(\omega)}\approx 0.7$, leads to a considerable improvement of the correspondence  of the experimental results
with calculations. 
It is clear, why the addition of a nonzero value of $\Re{\delta\eps_{\rm b}(\omega)}$ considerably changes the reflectivity 
under the conditions of the experiments discussed above, especially for smaller plasma densities and shorter wavelengths: for 
the considered plasma and laser radiation parameters the value of $\omega_{\rm pl}/\omega$ is not far from 1, 
being $\omega_{\rm pl}/\omega > 1$ and hence $\Re{\eps(\omega)} < 0$ for $n_{\rm i} > 10^{21}$ cm$^{-3}$. The addition of a small positive nonzero value for $\Re{\delta\eps_{\rm b}(\omega)}$ brings $\Re{\eps(\omega)}$ closer to zero, thus making the plasma more transparent for laser radiation. 

\begin{figure}
\includegraphics[width=0.9\columnwidth]{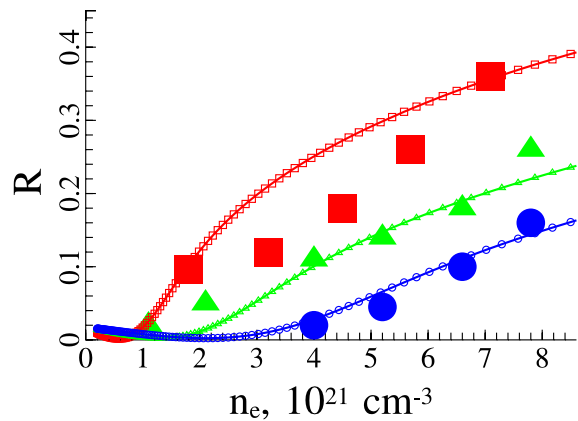}
\caption{(Color online).
 The same as in\FF{Xe_R_ro}, but with the account of a nonzero value of $\Re{\delta\eps_{\rm b}(\omega)}$ in the calculations (solid marked lines). A constant value $\Re{\delta\eps_{\rm b}(\omega)}=0.7$ was used in all calculations. 
}
\label{f:Xe_R_ro2}
\end{figure}





\section{Conclusions}\label{s:conclusions}


The LRT model has been used to develop a model for the intraband part of the permittivity of WDM. It is suitable in a wide frequency range, from the far infrared including the dc limit, till the X-ray spectral region. The model accounts for both electron-ion and electron-electron collisions, arbitrary degeneracy, screening and correlation effects. The relevant formulas are~\rf{Drude},\rf{nurw},\rf{rw},\rf{Rnm}--({\ref{Inm}),\rf{scr_dyn},\rf{Cnmdyn} and \rf{kappaD2}--(\ref{kappaD3}). Approximate expressions have been derived as simple fits that makes it possible to use them in hydrodynamic codes.

It is shown that the approach elaborated from the LRT is in good agreement with different variants of kinetic models derived on the base of the Boltzmann kinetic equation,  if higher order moments of the electron distribution function are taken into account. This holds for low frequencies ($\omega < \omega_{\rm pl}$) and for moderate coupling. 

At high frequencies ($\omega > \omega_{\rm pl}$) the introduction of  Dawson-Oberman-like corrections into classical kinetic models ensures a good agreement of 
the real part of the Drude-like effective collision frequency $\nu_{\rm Dr}$\rf{nuDr}, obtained within KT, with the results of LRT calculations. Nevertheless, the imaginary part of $\nu_{\rm Dr}$ is not correctly described by this correction procedure within the classical kinetic approach, and the LRT model should be used for description of $\Im{\nu_{\rm Dr}}$ in this frequency range. In addition, LRT gives a proper description of the inverse bremsstrahlung absorption of high-frequency laser radiation.

Simple expressions are obtained from the LRT approach in the region of low coupling where the Born approximation can be applied. Strong collisions are included within the Gould-DeWitt scheme results. The real and the imaginary part of permittivity calculated by the LRT model for optical frequencies and weak or moderate coupling are almost identical with those obtained within the kinetic approach.

Effects of screening are studied and it is shown, that statical screening of the interaction potential gives the same results as obtained from dynamical screening at any frequencies except the region of plasmon resonance in the vicinity of $\omega = \omega_{\rm pl}$. In the region of strong coupling, the limitation of the screening length by the interatomic distance is necessary for the correct description of optical properties of matter. Furthermore, ion correlations should be taken into account in the region of moderate and strong coupling for the accurate description of permittivity. 

As an application of the theory for calculating the DF,  optical properties of shock compressed noble gas plasmas have been considered. In addition to the contribution of free electrons (intraband contribution), also the contribution of bound electrons (inter band contribution) to the permittivity and the final width $L_S$ of the shock wave front must taken into account.
Furthermore, it was shown that for the considered plasma densities and wavelengths, a final width $L_S \gtrsim 3 l_s$ (where $l_s$  is skin layer depth) can lead to a considerable increase of the absorption coefficient $A$ and, this way, to a  change of the shape of the reflectivity curve $R(n)$  (where $R=1-A$ is the reflection coefficient and $n$ is the concentration of electrons) from initially convex to a more concave one. This behavior is more close to the results seen in experiments. On the other hand, such values of $L_S$ are about 30 times larger than estimates made for the equilibrium shock wave front width, determined by ion-ion collisions \cite{Stupochenko65PMTF,Zeldovich_Raizer}. Therefore it is necessary to take into account relaxation processes (ionization and excitation) for a more precise estimate for the width of a shock wave front. 
Our approach shows the possibility  to use optical measurements of the reflection coefficient of shock compressed gases as a tool for diagnostics of the structure of the shock wave front and relaxation processes in it. 

\section*{Acknowledgments}

We are grateful to N.E. Andreev for valuable discussions. M. Veysman thanks his German colleagues for hospitality and financial support during his visits to Rostock University and  Johannes Kepler University in Linz. 
The work of M. Veysman in JIHT was partly supported by the Presidium of
RAS program ``Thermo-physics of High Energy Density``, No I.13P.  
The authors acknowledge financial support by DFG  for  collaborative projects funded jointly with RAS. H.R.,  G.R. have also been supported by the DFG within the CRC-SFB 652.

\begin{appendix}

\section{Generalized linear response theory}\label{app1}

According to~\cite{Reinholz12PRE} (see also~\cite{ZMR97,ReinholzADP,Roepke88PRA}), 
the non-equilibrium statistical operator $\hat \rho(t)$ is determined by the dynamical evolution 
of the system with Hamiltonian $\hat H_{\rm tot} = \hat H + \hat H_{\rm ext}(t)$:
\begin{equation}
\hat \rho(t)=\lim_{\delta\to + 0} \delta \int\nolimits_{-\infty}^t dt' e^{-\delta (t-t')}\hat U(t,t') \hat \rho_{\rm rel}(t')\hat U^{\dag}(t,t') ,
\label{rho}
\end{equation}
where $\hat U(t,t')$ is the time evolution operator, which solves the  equation $i\hbar \partial_t\hat U(t,t') = \hat H_{\rm tot} \hat U(t,t') $
with initial condition  $\hat U(t',t')=1$. The system Hamiltonian  $\hat H$ is determined by Eq.\rf{H}. The external perturbation $\hat H_{\rm ext}(t)$ is determined in dipole approximation as 
\begin{equation}
\hat H_{\rm ext}(t) = -e \hat{\bs{R}}  \cdot \bs{E}(t), \; \hat{\bs{R}}= \sum\nolimits_i 
\hat{\bs{r}}_i, \; \hat{\dot{\bs{R}}} = \hat{\bs{P}}_{0,1}/m,
\label{Hext}
\end{equation}
$\hat{\bs{P}}_{0,1}$ is defined by Eq.\rf{Bn}. 

$\hat\rho_{\rm rel}(t)$ is the relevant statistical operator. It is introduced as a generalized Gibbs ensemble, which is derived from the principle of maximum of entropy:
\begin{eqnarray}
&& \hat\rho_{\rm rel}(t)= Z_{\rm rel}(t)^{-1} \exp\left[-\beta(\hat H-\mu \hat N)+\sum\nolimits_n F_n(t) \hat B_n\right], 
\nonumber\\ &&	
Z_{\rm rel}(t)= \Tr\left[-\beta(\hat H-\mu \hat N)+\sum\nolimits_n F_n(t) \hat B_n\right],
\label{rhorel}
\end{eqnarray}
where the Lagrange parameters $\beta$, $\mu$ and $F_n(t)$ are introduced to fix the given averages:
\begin{equation}
\Tr\left\{\hat B_n \rho(t)\right\}
= \langle \hat B_n^t \rangle = \Tr\left\{\hat B_n \rho_{\rm rel}(t)\right\},
\label{self-cons}
\end{equation} 
and similar equations holds for determination of $\beta$ and $\mu$ from conditions on 
$\langle \hat H \rangle$ and $\langle \hat N \rangle$, where $\hat N = \sum\nolimits_p \hat a_p^\dag \hat a_p$.
Equation\rf{rho} means that further correlations are build up from the initial state, determined by the relevant statistical operator $\hat\rho_{\rm rel}(t)$, and\rf{self-cons} means that observed statistical averages $\langle ..^t \rangle$ at time $t$ are correctly reproduced by $\hat \rho_{\rm rel} (t)$.

In LRT the response parameters $F_n(t)$ are considered to be small.  This permits one to perform the expansion of the relevant $\hat\rho_{\rm rel}(t)$ and the irrelevant $\hat\rho_{\rm irrel}(t)=\hat\rho(t)-\hat\rho_{\rm rel}(t) $ statistical operators with respect to $F_n(t)$, see~\cite{Reinholz12PRE}. Together with\rf{self-cons} and using the Kubo identity and partial integration of correlation functions, this give rise to the following system of equations: 
\begin{equation}
\langle\delta \hat B_n\rangle =\sum\nolimits_m 
( \hat B_n ;\delta \hat B_m) F_m,
\label{dBnav}
\end{equation}
\begin{multline}
\sum\limits_m \Bigl[ -i\omega\left\{
(\hat B_n;B_m) + \langle \hat{\dot{B}}_m; \delta B_m\rangle_z
\right\} + (\hat B_n;\hat{\dot{B}}_m) \\
+\langle \hat{\dot{B}}_n;\hat{\dot{B}}_m\rangle_z
\Bigr]F_m 
= \beta \frac em \left\{
(\hat B_n;\hat{\bs{P}}_1) + \langle \hat{\dot{B}}_m;\hat{\bs{P}}_1\rangle_z
\right\}\bs{E},
\label{Fmfull}
\end{multline}
where $z=\omega+i\delta$;  
 $\delta \hat B_n = \hat B_n - \langle \hat B_n \rangle_0$,
$\langle \hat B_n \rangle_0$ is the statistical average of $\hat B_n $ with the equilibrium density operator $\rho_0$.

The quantity $\hat{\bs{P}}_1 = \hat{\bs{B}_0}(0)=\hat{\bs{P}}_{0,1}$ is the operator of the total momentum of electrons given by Eq.\rf{Bn} (we consider the long-wavelength limit $k\to 0$).
The operators $\hat{{B}}_n$ are also chosen in the form of Eq.\rf{Bn} (as well as $\hat{\bs{P}}_1$, they are vectors). 

At the leading order of the parameter of interaction, proportional to $e^2$, one can show~\cite{Reinholz00PRE} that the terms containing only one operator $\hat{\dot{B}}_n$ can be omitted. For the  set \rf{Bn}  of observables, the equilibrium averages vanish, $\langle \hat B_n \rangle_0 =0$. Taking this in mind, one can derive from\rf{dBnav} and\rf{Fmfull} the following expressions, which determine the values of the density of electric current $\langle \hat{\bs{J}}\rangle = e \langle \hat{\bs{P}}_1\rangle$ and the response parameters $F_n$:

\begin{equation}
\bs{J} = \frac{n e^2}{m} \bs{E}\sum\nolimits_m \Nf_{1m} \F_m ,
\label{J}
\end{equation}

\begin{equation}
\sum\limits_m \Bigl[ \Cf_{nm}-i\w \Nf_{nm}\Bigr]\F_m 
= \Nf_{n1},
\label{Fm}
\end{equation}

where the dimensionless correlation functions and response parameters are defined in Eq. (\ref{dimcorr}).

\section{Evaluation of  correlation functions}\label{app2}

Correlation functions introduced in Eq.\rf{dimcorr} can be expressed as
\begin{multline}\label{Cnm2ei}
\Cf_{nm}^{\rm ei}(\omega)=i Z/(3\pi^2) \\
 \times \int_0^\infty  f_{\rm scr}(y) dy \int_{-\infty}^\infty \frac{dx}{x} 
\frac{R_{nm}^{\rm ei}(x,y)}{w+i \delta -x } 
\ln \left[\frac{1+e^{\emu-(x/y-y)^2}}{1+e^{\emu-(x/y+y)^2}}\right],
\end{multline}
\begin{multline}\label{Cnm2ee}
\Cf_{nm}^{\rm ee}(\omega)=i /(3\sqrt{2}\pi^2) \\ 
 \times
\int_0^\infty  f_{\rm scr}^{\rm e} (y) dy \int_{-\infty}^\infty \frac{dx}{x} 
\frac{R_{nm}^{\rm ee}(x,y)}{w+i \delta -x } 
\ln \left[\frac{1+e^{\emu-(x/y-y)^2}}{1+e^{\emu-(x/y+y)^2}}\right],
\end{multline}
where $\Cf_{nm}^{\rm ee}(\omega)$ and $\Cf_{nm}^{\rm ei}(\omega)$  are the contributions owing to electron-electron and  
 electron-ion interaction, respectively, and 
$\Cf_{nm}=\Cf_{nm}^{\rm ee}+\Cf_{nm}^{\rm ei}$.
Expressions $R_{nm}^{\rm ei}$ and $R_{nm}^{\rm ee}$ are polynomials of $x$ and $y$. For $n,m=1,3$ they have the following form, see~\cite{Reinholz12PRE}:
\begin{eqnarray}
&&R_{11}^{\rm ei}=1,\; R_{31}^{\rm ei}=R_{13}^{\rm ei}=1+y^2+3x^2,\;  
\nonumber \\ &&
R_{33}^{\rm ei}=2+2y^2+y^4 + 2x^2(5+3y^2)+9x^4,
\nonumber \\ &&	
R_{11}^{\rm ee}= R_{31}^{\rm ee}=R_{13}^{\rm ee}=0,\; R_{33}^{\rm ee}=1+19 x^2/4. 
\label{Rnm}
\end{eqnarray}
Similar expressions can be given for the higher order polynoms, see Refs. \cite{Redmer97PhysRep,Karakhtanov13CTPP}.

The screening function $f_{\rm scr}^{\rm e}(y)$ is defined as
\begin{equation}f_{\rm scr}^{\rm e}(y) = y^3/[y^2 + \k_{\rm D}^2/4]
\label{fscre}
\end{equation}
and the screening function $f_{\rm scr}^{\rm i}(y)\equiv f_{\rm scr}(y)$ is defined above, see Eqs.\rf{fscr}, \rf{scr_dyn}, \rf{Cnmdyn} and Eq.\rf{fGericke} in the case of  pseudopotentials. The value of $\k_{\rm D}$ is given by Eq.\rf{kappaD2}.
Note that in\rf{fscre} $\k_{\rm D}$ contains the numerical factor $1/4$ instead of $1/8$ in the similar expression\rf{fscr}.

As done above, see Eqs.\rf{nuBsc2Re} and\rf{nuBsc2Im}, the correlation functions can be decomposed into a real part and a imaginary part using the Sokhotski-Plemej formula. One  obtains for the real part of the correlation functions the expression 
\begin{multline}\label{Cnmr}
{\Cf'}_{nm}^{{\rm e}q}= \alpha_q/(3\pi w)
 \\ \times 
\int_0^\infty  f_{\rm scr}^{q}(y) dy R_{nm}^{{\rm e}q}\left(\frac wy,y\right)
\ln \left[\frac{1+e^{\emu -(w/y-y)^2}}{1+e^{\emu-(w/y+y)^2}}\right],
\end{multline}
where $q={\rm i}$ or $\rm e$, $\alpha_{\rm i}=Z$, $\alpha_{\rm e}=1/\sqrt{2}$.

For the imaginary part of the correlation functions one obtains 
\begin{equation}\label{Cnmi}
{\Cf''}_{nm}^{{\rm e}q}= \dst\frac{\alpha_q}{3\pi^2 w}\int\limits_0^\infty  \!f_{\rm scr}^{q}(y) dy\left[\sum\limits_{\delta=\pm 1} \I_{nm}^{{\rm e}q,\delta} (y)-
2\I_{nm}^{{\rm e}q,0}(y)\right], 
\end{equation}
\begin{eqnarray}
&&\I_{nm}^{e q,l}=
\int\limits_0^\infty \dst\frac{d\xi}{\xi}\sum\limits_{\sigma=\pm 1}\sigma
R_{nm}^{{\rm e} q}\left(\xi+\sigma\frac{lw}{y},y\right) \nonumber \\ &&
\times \ln\left[
1+e^{\emu -[\xi+\sigma (y + lw/y)]^2}\right],
\label{Inm}
\end{eqnarray}
with $l=0,\pm 1$.

\end{appendix}




\bibliography{QS_2014_56}

\end{document}